%% file: main.tex
\documentclass[submission,copyright,creativecommons]{eptcs}
\providecommand{\event}{QPL 2022} %
\usepackage{underscore}           %

\usepackage{amsmath}
\usepackage{amssymb}
\usepackage{braket}
\usepackage{fancyhdr}
\usepackage{booktabs}

\usepackage[T1]{fontenc}
\usepackage[utf8]{inputenc}
\usepackage{regexpatch}
\usepackage{letltxmacro}
\usepackage{aligned-overset}

\usepackage{hyperref}
\usepackage[noabbrev,nameinlink]{cleveref}

\usepackage{enumitem}

\usepackage{xspace}

\usepackage{nowidow}
\usepackage{microtype}

\usepackage{float}
\usepackage{caption}
\usepackage{subcaption}

\usepackage{xcolor}

\usepackage{cancel}

\usepackage{algorithm}
\usepackage[noend]{algpseudocode}

\usepackage{wrapfig}

\usepackage{svg}

\usepackage{tikz}

\usetikzlibrary{backgrounds}
\usetikzlibrary{arrows}
\usetikzlibrary{shapes,shapes.geometric,shapes.misc}
\usetikzlibrary{decorations.pathmorphing}
\usetikzlibrary{decorations.pathreplacing}
\usetikzlibrary{decorations.markings}

\tikzstyle{every picture}=[baseline=-0.25em,scale=0.5]

\pgfkeys{/tikz/tikzit fill/.initial=0}
\pgfkeys{/tikz/tikzit draw/.initial=0}
\pgfkeys{/tikz/tikzit shape/.initial=0}
\pgfkeys{/tikz/tikzit category/.initial=0}

\newcommand{\tikzfig}[1]{%
\IfFileExists{#1.tikz}
  {\input{#1.tikz}}
  {%
    \IfFileExists{./figures/#1.tikz}
      {\input{./figures/#1.tikz}}
      {\tikz[baseline=-0.5em]{\node[draw=red,font=\color{red},fill=red!10!white] {\textit{#1}};}}%
  }%
}

\newcommand{\oftype}[2]{\text{#1}\,:\,\text{#2}}
\newcommand{\ZX}[2]{\texttt{ZX}\,\,\text{#1}\,\,\text{#2}}

\newcommand{\VyZX}{\textsl{Vy}\textsc{ZX}\xspace}
\newcommand{\SQIR}{\textsc{sqir}\xspace}
\newcommand{\QASM}{\texttt{QASM}\xspace}
\newcommand{\QLib}{\texttt{QuantumLib}\xspace}
\newcommand{\pyZX}{PyZX\xspace}
\newcommand{\VOQC}{\textsc{Voqc}\xspace}

\newcommand{\certiq}{\texttt{CertiQ}\xspace}
\newcommand{\tket}{\texttt{t\(\ket{\text{ket}}\)}\xspace}
\newcommand{\quartz}{Quartz\xspace}

\hypersetup{
    colorlinks,
    linkcolor={red!50!black},
    citecolor={blue!50!black},
    urlcolor={blue!80!black}
}

\usepackage{mathpartir}
\usepackage{etoolbox}

\definecolor{egreen}{rgb}{0.31, 0.78, 0.47}
\definecolor{ate}{rgb}{0.58, 0.0, 0.83}

\newcommand{\R}{\mathbb{R}}
\newcommand{\C}{\mathbb{C}}
\newcommand{\N}{\mathbb{N}}

\pgfdeclarelayer{edgelayer}
\pgfdeclarelayer{nodelayer}
\pgfsetlayers{background, edgelayer, nodelayer, main}
\tikzstyle{none}=[inner sep=0mm]
\tikzstyle{every loop}=[]
\tikzstyle{mark coordinate}=[inner sep=0pt,outer sep=0pt,minimum size=3pt,fill=black,circle]
\input{zh.tikzdefs}
\input{zh.tikzstyles}

\algnewcommand\algorithmicswitch{\textbf{switch}}
\algnewcommand\algorithmiccase{\textbf{case}}
\algnewcommand\algorithmicassert{\texttt{assert}}

\algdef{SE}[SWITCH]{Switch}{EndSwitch}[1]{\algorithmicswitch\ #1\ \algorithmicdo}{\algorithmicend\ \algorithmicswitch}%
\algdef{SE}[CASE]{Case}{EndCase}[1]{\algorithmiccase\ #1}{\algorithmicend\ \algorithmiccase}%
\algtext*{EndSwitch}%
\algtext*{EndCase}%

\makeatletter
\def\blfootnote{\gdef\@thefnmark{}\@footnotetext}
\makeatother

\newcommand{\mailtodomain}[1]{\href{mailto:#1}{\nolinkurl{#1}}}

\newcommand{\titletext}{\VyZX : A Vision for Verifying the ZX Calculus}

\title{\titletext}
\author{
Adrian Lehmann*
\institute{University of Chicago}
\email{\mailtodomain{adrianlehmann@uchicago.edu}}
\and
Ben Caldwell*
\institute{University of Chicago}
\email{\mailtodomain{caldwellb@uchicago.edu}}
\and
Robert Rand
\institute{University of Chicago}
\email{\mailtodomain{rand@uchicago.edu}}
}

\def\titlerunning{\titletext}
\def\authorrunning{A. Lehmann, B. Caldwell \& R. Rand}
\begin{document}
\maketitle
\blfootnote{* Equal contribution}

\input{sections/abstract}

\input{sections/introduction}

\input{sections/prior-work}

\input{sections/vyzx}

\input{sections/future}

\input{sections/conclusion}

\section*{Acknowledgements}
This material is based upon work supported by EPiQC, an NSF Expedition in Computing,
under Grant No. CCF-1730449 and the Air Force Office of Scientific Research under Grant No.
FA95502110051.

\nocite{*}
\bibliographystyle{eptcs}
\bibliography{generic}

\end{document}

%% file: zh.tikzstyles
\tikzstyle{dot}=[inner sep=0.3mm, minimum width=2mm, minimum height=2mm, draw, shape=circle, font={\footnotesize}, tikzit fill=magenta]
\tikzstyle{white dot}=[dot, fill={rgb,255: red,232; green,165; blue,165}, text depth=-0.2mm, tikzit category=ZH-pf, draw=black]
\tikzstyle{white phase dot}=[minimum size=5mm, font={\footnotesize\boldmath}, shape=rectangle, rounded corners=2mm, inner sep=0.2mm, outer sep=-2mm, scale=0.8, tikzit shape=circle, draw=black, fill={rgb,255: red,232; green,165; blue,165}, tikzit category=ZH-pf, tikzit draw=blue]
\tikzstyle{gray dot}=[dot, fill={rgb,255: red,216; green,248; blue,216}, text depth=-0.2mm, tikzit category=ZH-pf]
\tikzstyle{gray phase dot}=[white phase dot, tikzit shape=circle, tikzit draw=blue, fill={rgb,255: red,216; green,248; blue,216}, font={\footnotesize\boldmath}]
\tikzstyle{hadamard}=[fill=white, draw, inner sep=0.6mm, minimum height=1.5mm, minimum width=1.5mm, shape=rectangle, tikzit shape=rectangle, tikzit category=ZH-pf]
\tikzstyle{small hadamard}=[hadamard]
\tikzstyle{lambda}=[hadamard, fill={rgb,255: red,180; green,180; blue,180}, tikzit shape=rectangle]
\tikzstyle{halfscalar}=[star, fill=black, draw=black, minimum size=8pt, inner sep=0pt]
\tikzstyle{box}=[shape=rectangle, text height=1.5ex, text depth=0.25ex, yshift=0.2mm, fill=white, draw=black, minimum height=3mm, minimum width=5mm, font={\small}]
\tikzstyle{Z dot}=[inner sep=0mm, minimum size=2mm, shape=circle, draw=black, fill={zx_green}, tikzit fill=green]
\tikzstyle{Z phase dot}=[minimum size=5mm, font={\footnotesize\boldmath}, shape=rectangle, rounded corners=2mm, inner sep=0.2mm, outer sep=-2mm, scale=0.8, tikzit shape=circle, draw=black, fill={zx_green}, tikzit draw=blue, tikzit fill=green]
\tikzstyle{X dot}=[Z dot, shape=circle, draw=black, fill={zx_red}, tikzit fill=red]
\tikzstyle{X phase dot}=[Z phase dot, tikzit shape=circle, tikzit draw=blue, fill={zx_red}, font={\footnotesize\color{black}\boldmath}, tikzit fill=red]
\tikzstyle{H box}=[hadamard]
\tikzstyle{st}=[star, star points=5, fill=white, draw=black, inner sep=1.2pt, line width=1.2pt, tikzit fill=blue, tikzit draw=red, tikzit category=ZH-pf]
\tikzstyle{triangle}=[regular polygon, regular polygon sides=3, fill=white, draw=black, inner sep=0pt, minimum width=1em, tikzit draw=blue, tikzit category=ZH-pf, tikzit fill=cyan]
\tikzstyle{not}=[fill={rgb,255: red,180; green,180; blue,180}, draw=black, shape=circle, font={$\neg$}, dot]
\tikzstyle{vertex}=[inner sep=0mm, minimum size=1mm, shape=circle, draw=black, fill=black]
\tikzstyle{vertex set}=[inner sep=0mm, minimum size=1mm, shape=circle, draw=black, fill=white, font={\footnotesize\boldmath}]
\tikzstyle{wide point}=[fill=white, draw, shape=isosceles triangle, shape border rotate=-90, isosceles triangle stretches=true, inner sep=0pt, minimum width=1.5cm, minimum height=6.12mm, yshift=-0.0mm]
\tikzstyle{medium gray box}=[semilarge box, fill={rgb,255: red,180; green,180; blue,180}]
\tikzstyle{small box}=[rectangle, inline text, fill=white, draw, minimum height=5mm, yshift=-0.5mm, minimum width=5mm, font={\small}]
\tikzstyle{small gray box}=[small box, fill={rgb,255: red,180; green,180; blue,180}]
\tikzstyle{medium box}=[rectangle, inline text, fill=white, draw, minimum height=5mm, yshift=-0.5mm, minimum width=8mm, font={\small}]
\tikzstyle{ddot}=[line width=1.6pt, inner sep=0mm, minimum width=2.5mm, minimum height=2.5mm, draw, shape=circle]
\tikzstyle{dd white}=[ddot, fill=white, tikzit draw=green]
\tikzstyle{dd white phase}=[white phase dot, line width=1.6pt, tikzit draw=yellow]
\tikzstyle{dd gray}=[ddot, fill={rgb,255: red,180; green,180; blue,180}, tikzit draw=green]
\tikzstyle{dd gray phase}=[gray phase dot, line width=1.6pt, tikzit draw=yellow]

\tikzstyle{simple}=[-, fill=none]
\tikzstyle{hadamard edge}=[-, dashed, dash pattern=on 2pt off 1pt, thick, draw=gray]
\tikzstyle{gray}=[-, draw={blue!60!white}, tikzit draw=blue]
\tikzstyle{blue}=[-, draw={blue!60!white}, tikzit draw=blue]
\tikzstyle{brace edge}=[-, tikzit draw=blue, decorate, decoration={brace,amplitude=1mm,raise=-1mm}]
\tikzstyle{diredge}=[->]
\tikzstyle{not edge}=[-, dashed, dash pattern=on 2pt off 1.5pt, thick, draw={rgb,255: red,255; green,68; blue,68}]
\tikzstyle{double edge}=[-, double, shorten <=-1mm, shorten >=-1mm, double distance=2pt]
\tikzstyle{boldedge}=[-, line width=1.6pt, shorten <=-0.17mm, shorten >=-0.17mm, tikzit draw=blue]
\tikzstyle{separatediagrams}=[-, dashed, dash pattern=on 2pt off 1.5pt, thick, draw=black]

%% file: sections/abstract.tex
\begin{abstract}
Optimizing quantum circuits is a key challenge for quantum computing. The PyZX compiler broke new ground by optimizing circuits via the ZX calculus, a powerful graphical alternative to the quantum circuit model. Still, it carries no guarantee of its correctness.
To address this, we developed \VyZX, a verified ZX-calculus in the Coq proof assistant.
\VyZX provides two distinct representations of ZX diagrams for ease of programming and proof: A graph-based representation for writing high-level functions on diagrams and a block-based representation for proving ZX diagrams equivalent. 
Through these two different views, \VyZX provides the tools necessary to verify properties and transformations of ZX diagrams.
This paper explores the proofs and design choices underlying \VyZX and its application and the challenges of verifying a graphical programming language.
\end{abstract}

%% file: sections/introduction.tex
\section{Introduction}

As quantum computers transition from fiction to a feature of our daily lives, there has been a surge of interest in quantum optimizers~\cite{nam2018automated,shi2019certiq,smith2019quantum,kissinger2020Pyzx,Amy2019,mingkuan2022quartz}. The goal of a quantum optimizer is to reduce the number of bottlenecks in a quantum circuit, whether those be two-qubit gates in the near term or $T$ gates in the longer term. Many of these optimizers do some form of model checking~\cite{shi2019certiq,mingkuan2022quartz} or translation validation~\cite{smith2019quantum,kissinger2020Pyzx} to gain confidence that their optimizations are correct, out of awareness that bugs in quantum optimizers are both common and costly~\cite{kissinger2020Pyzx}. Of particular note, the \VOQC compiler~\cite{hietala-et-al-2021-VOQC} is fully verified in the Coq proof assistant, guaranteeing that its optimizations preserve the semantics of the original quantum circuit.

Unfortunately, the quantum circuit model has many weaknesses, particularly from an optimization perspective. 
Quantum circuits come equipped with a variety of different gate sets: A good optimizer for the Clifford+T gate set is not guaranteed to perform well on IBM's gates or Google's. 
They are also rigid: They consist of a large sequence of vertically and horizontally ordered gates, whereas an optimizer only cares about the connections between gates. 
In this spirit, Kissinger and van de Wetering developed PyZX~\cite{kissinger2020Pyzx}, an optimizer for the ZX calculus~\cite{coecke-duncan-zx}, a graphical language for quantum computing in which \emph{only connectivity matters}~\cite{coecke-kissinger-2017-picturing-q-proc}. 
Like prior tools \pyZX checks for correctness by translation validation, either converting diagrams to their underlying linear maps in NumPy and checking if all elements of the linear map are equal up to a global nonzero scalar, or ``optimizing'' a circuit concatenated with its adjoint, and checking to see if it returns the identity.
Unfortunately, these methods are slow and are not guaranteed to succeed as showing circuit equivalence is known to be QMA-complete in the general case~\cite{janzing2003-id-qma}.

Drawing inspiration from \VOQC and \pyZX and the verified classical compiler CompCert~\cite{leroy2009compcert}, we present \VyZX, a formalization of the ZX calculus in the Coq~\cite{Coq12} proof assistant. \VyZX is intended to be a fully verified implementation of the \pyZX compiler and a platform for mechanized reasoning about the ZX calculus and related graphical calculi. Given the versatility of the ZX-calculus, \VyZX should allow us to tackle correctness issues in a range of domains, including lattice surgery~\cite{deBeaudrap2020zxcalculusis}, circuit simulation~\cite{kissinger2020simulation}, and natural language processing~\cite{coecke2020nlpzx}.

Unfortunately, while ``only connectivity matters'' is an excellent slogan for a graphical language, it poses significant challenges for formal verification. 
Computers do not talk in pictures: Internally, they impose a rigidity akin to that of the circuit model. Even the standard representations of graphs, like adjacency lists and adjacency matrices, are ill-suited to inductive reasoning of the sort Coq excels in. 
To address this, we formalize two views of ZX diagrams: \emph{block representation}, in which wires and nodes are composed horizontally and vertically, and \emph{graph representation}, which is more faithful to the standard representation of ZX diagrams.

We explore the motivation for creating \VyZX in \Cref{sec:prior}. We layout the design decisions underlying \VyZX and discuss their potential in \Cref{sec:vyzx}. There we also cover the inductive definition for ZX diagrams, how we apply semantics to them, and how we prove equivalence of two diagrams. We discuss how to convert from standard quantum circuits to our inductive diagrams (\Cref{sec:ingest}) and how we can view our inductive diagrams as graphs (\Cref{sec:graph}). We sketch out a path from our current formalization of the ZX calculus to a full-fledged quantum optimizer that is integrated with the \VOQC compiler and conclude with the many potential use-cases for \VyZX in \Cref{sec:future}. 
All code referenced in this paper can be found at \url{https://github.com/inQWIRE/VyZX}.

%% file: sections/prior-work.tex
\section{Verified Optimization and the ZX-Calculus}\label{sec:prior}

\subsection{Verified Optimization}

Quantum circuit optimizers take varied approaches to verify that their optimizations are well behaved.
Some compilers, such as the one created by Nam et al.~\cite{nam2018automated}, solely rely on unit testing to ensure correctness, though unit testing can only show the presence -- not absence -- of bugs.
Quantinuum's \tket~\cite{Sivarajah2020tket} goes a step further and uses a Hoare logic system to check that certain postconditions, such as ``the circuit contains no wire swaps'', hold for various optimization passes.
This is a useful system for a quantum optimizer to have, but %
checking that the postcondition holds does not guarantee that the optimization has returned an equivalent circuit to the input circuit.
This gap in how we can verify quantum circuit optimizations has been filled by a few other compilers.

The quantum compilers \certiq~\cite{shi2019certiq} and \quartz~\cite{mingkuan2022quartz} attempt to alleviate this concern by adding systems to check for circuit equivalence.
They use these systems throughout their development process to check optimizations.
To check equivalence, they generate some proof obligations as SMT formulas and pass them along to Z3.
This steps beyond compilers like \tket as it actually attempts to validate the optimization.
The key feature of compilers like \quartz and \certiq is that they automate this validation of optimization passes.
While this is a valuable way to make it easy to engineer new optimization passes, it is incomplete.
Not every optimization can be verified by an SMT solver, and compilers using such solvers hence include certain optimization without validation.
This undermines the validation itself if the entire thing is not validated.
If we want to verify the optimizer completely, we need a stronger system that allows us to prove the correctness, not just pass it to an SMT solver.

The flaws of existing compilers inspired the development of the \VOQC verified compiler~\cite{hietala-et-al-2021-VOQC}.
\VOQC uses \SQIR~\cite{hietala-et-al:LIPIcs.ITP.2021.21} and \QLib~\cite{QuantumLib} to provide a full-stack verification pipeline.
All three libraries above are written in the Coq proofs assistant, providing them with strong correctness guarantees.
\VOQC ingests \SQIR circuits from \QASM~\cite{cross2017openqasm}, then applies optimization passes that are proven correct at compile time in Coq, and hence do not require any automation as we have seen with \certiq.
Upon completion \SQIR circuits are converted back into \QASM.
\SQIR can also handle multiple different gate sets, which are related to a base gate set that is used for proof.

\subsection{ZX calculus}\label{sec:zx}

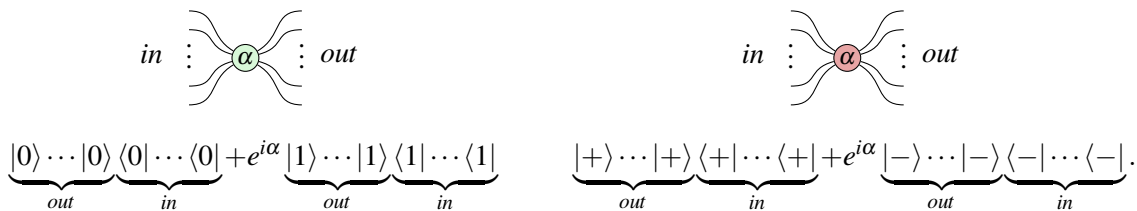
\begin{figure}[H]
    \centering
    \begin{subfigure}{.5\textwidth}
        \centering
        \input{Z-Spider.tikz}
    \end{subfigure}\begin{subfigure}{.5\textwidth}
        \centering
        \input{X-Spider.tikz}
    \end{subfigure}
    
    \par\bigskip
    
    \begin{subfigure}{.5\textwidth}
    \centering
    \(\underbrace{\ket{0}\cdots\ket{0}}_{out}  \underbrace{\bra{0}\cdots\bra{0}}_{in} + e^{i\alpha} \underbrace{\ket{1}\cdots\ket{1}}_{out} \underbrace{\bra{1}\cdots\bra{1}}_{in}\)
    \end{subfigure}\begin{subfigure}{.5\textwidth}
    \centering
    \(
    \underbrace{\ket{+}\cdots\ket{+}}_{out}  \underbrace{\bra{+}\cdots\bra{+}}_{in} + e^{i\alpha} \underbrace{\ket{-}\cdots\ket{-}}_{out} \underbrace{\bra{-}\cdots\bra{-}}_{in}.
    \)
    \end{subfigure}
    \caption{\centering Z and X spiders with their standard bra-ket semantics.}\label{fig:eqnXZ}
    \vspace*{-4mm}
\end{figure}

In contrast to the circuit model used by \VOQC and other optimizers, ZX diagrams are a graphical representation of quantum operations.
The ZX calculus~\cite{CoeckeDuncan2011} uses such diagrams together with a set of rewrite rules to manipulate quantum operations.
Fundamentally, ZX diagrams are graphs with green and red nodes\footnote{We chose accessible shades of green and red for this paper (see \url{https://zxcalculus.com/accessibility.html})}, called Z and X spiders, with $in$ inputs and $out$ outputs, along with a rotation angle $\alpha \in [0,2\pi)$. If the rotation angle is $0$ it can be omitted. The semantics of Z spiders and X spiders are shown in \Cref{fig:eqnXZ}.

These spiders are connected through edges.
Edges can either be regular edges or Hadamard edges, which are represented as dotted lines in diagrams and implicitly add a Hadamard gate on their path. These Hadamard edges can be treated as syntax for regular edges with three nodes, see \Cref{fig:gates}.
For more on the ZX-calculus, we recommend John van de Wetering's excellent survey of the topic~\cite{vandewetering2020zxcalculus}.

\subsection{ZX Calculus Optimization}\label{sec:zx-opt}

Our work is inspired by \pyZX optimizations, which is based on work by Duncan et al.~\cite{duncan-et-al-2020} and Kissinger and van de Wetering~\cite{kissinger-wetering-2020}.
Duncan et al.'s describe how to optimize ZX diagrams with graph-theoretic rules.
They begin by outlining a restriction on ZX diagrams, called \textit{graph-like form}.
Such diagrams are restricted to having only one kind of node (the Z spider), one kind of edge (Hadamard edge), no self-loops, and the condition that each node has a unique parent (i.e., either only one input or output).
They show that one can freely convert between unrestricted ZX diagrams and graph-like ZX diagrams with equivalent semantics.
Using graph-like form, they use graph operations, such as local complementation and pivoting, to reduce the number of Clifford gates (corresponding to nodes containing multiples of $\pi/2$) in their circuits by combining nodes.

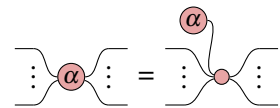
\begin{wrapfigure}[6]{r}{.25\textwidth}
    \centering
    \input{phase-gadget.tikz}
    \caption{\centering A phase gadget}\label{fig:phase-gadget}
\end{wrapfigure}

Since the algorithm above does not reduce the number of non-Clifford gates in a circuit, Kissinger and van de Wetering~\cite{kissinger-wetering-2020} devised an optimization strategy for ZX calculus to reduce the T-gate count, $T$ gates beings the most expensive operations for error-corrected quantum computers.
To achieve this, they remove any non-Clifford angles from spiders by splitting them into so-called \textit{phase gadgets}, as shown in \Cref{fig:phase-gadget}.
Then, the resulting Clifford gates are optimized as previously described.
Finally, phase gadgets are fused back into the diagram resulting in an optimized diagram.
In merging multiple phase gadgets spiders, the phase gadgets themselves can be merged, creating angles that are multiples of \(\frac{\pi}{2}\) that can be fused into the Clifford part of the diagram, reducing the number of non-Clifford gates.
With these optimizations, the ZX-based optimizer \pyZX achieves state-of-the-art performance~\cite{kissinger2020Pyzx} on Maslov's reversible benchmark suite~\cite{maslov2021benchmark}. 

\VOQC and \pyZX both stand out as significant quantum optimizers for their use of formal verification and the ZX-calculus, respectively. 
With \VyZX, we set out to combine these two ideas into one quantum circuit optimizer.

%% file: figures/Z-spider.tikz
\begin{tikzpicture}
	\begin{pgfonlayer}{nodelayer}
		\node [style=gray dot] (0) at (0, 0) {$\alpha$};
		\node [style=none] (1) at (1.5, 1.25) {};
		\node [style=none] (2) at (1.5, -1.25) {};
		\node [style=none] (3) at (-1.5, 1.25) {};
		\node [style=none] (4) at (-1.5, -1.25) {};
		\node [style=none] (5) at (1.5, 0) {};
		\node [style=none] (7) at (1.5, 0.25) {\vdots};
		\node [style=none] (8) at (-1.5, 0.25) {\vdots};
		\node [style=none] (9) at (-2.5, 0.1) {$in$};
		\node [style=none] (10) at (2.5, 0.1) {$out$};
		\node [style=none] (11) at (-1.5, 0.75) {};
		\node [style=none] (12) at (-1.5, -0.75) {};
		\node [style=none] (13) at (1.5, -0.75) {};
		\node [style=none] (14) at (1.5, 0.75) {};
	\end{pgfonlayer}
	\begin{pgfonlayer}{edgelayer}
		\draw [in=180, out=30, looseness=1.25] (0) to (1.center);
		\draw [in=180, out=-30, looseness=1.25] (0) to (2.center);
		\draw [in=150, out=0, looseness=1.25] (3.center) to (0);
		\draw [in=-150, out=0, looseness=1.25] (4.center) to (0);
		\draw [in=165, out=0, looseness=1.50] (11.center) to (0);
		\draw [in=-165, out=0, looseness=1.50] (12.center) to (0);
		\draw [in=-180, out=15, looseness=1.50] (0) to (14.center);
		\draw [in=-180, out=-15, looseness=1.50] (0) to (13.center);
	\end{pgfonlayer}
\end{tikzpicture}

%% file: figures/X-spider.tikz
\begin{tikzpicture}
	\begin{pgfonlayer}{nodelayer}
		\node [style=white dot] (0) at (0, 0) {$\alpha$};
		\node [style=none] (1) at (1.5, 1.25) {};
		\node [style=none] (2) at (1.5, -1.25) {};
		\node [style=none] (3) at (-1.5, 1.25) {};
		\node [style=none] (4) at (-1.5, -1.25) {};
		\node [style=none] (5) at (1.5, 0.25) {};
		\node [style=none] (6) at (1.5, 0.25) {};
		\node [style=none] (7) at (1.5, 0.25) {\vdots};
		\node [style=none] (8) at (-1.5, 0.25) {\vdots};
		\node [style=none] (9) at (-2.5, 0.1) {$in$};
		\node [style=none] (10) at (2.5, 0.1) {$out$};
		\node [style=none] (11) at (-1.5, 0.75) {};
		\node [style=none] (12) at (-1.5, -0.75) {};
		\node [style=none] (13) at (1.5, -0.75) {};
		\node [style=none] (14) at (1.5, 0.75) {};
	\end{pgfonlayer}
	\begin{pgfonlayer}{edgelayer}
		\draw [in=180, out=30, looseness=1.25] (0) to (1.center);
		\draw [in=180, out=-30, looseness=1.25] (0) to (2.center);
		\draw [in=150, out=0, looseness=1.25] (3.center) to (0);
		\draw [in=-150, out=0, looseness=1.25] (4.center) to (0);
		\draw [in=165, out=0, looseness=1.50] (11.center) to (0);
		\draw [in=-165, out=0, looseness=1.50] (12.center) to (0);
		\draw [in=-180, out=15, looseness=1.50] (0) to (14.center);
		\draw [in=-180, out=-15, looseness=1.50] (0) to (13.center);
	\end{pgfonlayer}
\end{tikzpicture}

%% file: figures/phase-gadget.tikz
\begin{tikzpicture}
	\begin{pgfonlayer}{nodelayer}
		\node [style=white dot] (18) at (6, 0) {$\alpha$};
		\node [style=none] (19) at (5, -0.75) {};
		\node [style=none] (20) at (5, 0.75) {};
		\node [style=none] (24) at (7, 0.75) {};
		\node [style=none] (25) at (7, -0.75) {};
		\node [style=none] (44) at (8, 0) {=};
		\node [style=white dot] (45) at (10, 0) {};
		\node [style=none] (46) at (9, -0.75) {};
		\node [style=none] (47) at (9, 0.75) {};
		\node [style=white dot] (48) at (9.25, 1.5) {$\alpha$};
		\node [style=none] (50) at (11, 0.75) {};
		\node [style=none] (51) at (11, -0.75) {};
		\node [style=none] (53) at (9, 0) {.};
		\node [style=none] (54) at (9, -0.25) {.};
		\node [style=none] (58) at (8.5, 0.75) {};
		\node [style=none] (59) at (8.5, -0.75) {};
		\node [style=none] (60) at (11.5, 0.75) {};
		\node [style=none] (61) at (11.5, -0.75) {};
		\node [style=none] (62) at (4.5, 0.75) {};
		\node [style=none] (63) at (4.5, -0.75) {};
		\node [style=none] (64) at (7.5, 0.75) {};
		\node [style=none] (65) at (7.5, -0.75) {};
		\node [style=none] (66) at (9, 0.25) {.};
		\node [style=none] (68) at (11, 0) {.};
		\node [style=none] (69) at (11, -0.25) {.};
		\node [style=none] (70) at (11, 0.25) {.};
		\node [style=none] (72) at (5, 0) {.};
		\node [style=none] (73) at (5, -0.25) {.};
		\node [style=none] (74) at (5, 0.25) {.};
		\node [style=none] (76) at (7, 0) {.};
		\node [style=none] (77) at (7, -0.25) {.};
		\node [style=none] (78) at (7, 0.25) {.};
		\node [style=none] (79) at (12, 0.5) {};
	\end{pgfonlayer}
	\begin{pgfonlayer}{edgelayer}
		\draw [style=simple, in=0, out=180] (24.center) to (18);
		\draw [style=simple, in=180, out=-15] (18) to (25.center);
		\draw [style=simple, in=0, out=-165] (18) to (19.center);
		\draw [style=simple, in=0, out=-180] (18) to (20.center);
		\draw [in=165, out=0, looseness=0.75] (48) to (45);
		\draw [style=simple, in=0, out=180] (50.center) to (45);
		\draw [style=simple, in=180, out=-15] (45) to (51.center);
		\draw [style=simple, in=0, out=-165] (45) to (46.center);
		\draw [style=simple, in=0, out=-180] (45) to (47.center);
		\draw (24.center) to (64.center);
		\draw (25.center) to (65.center);
		\draw (19.center) to (63.center);
		\draw (62.center) to (20.center);
		\draw (58.center) to (47.center);
		\draw (46.center) to (59.center);
		\draw (51.center) to (61.center);
		\draw (50.center) to (60.center);
	\end{pgfonlayer}
\end{tikzpicture}

%% file: sections/vyzx.tex
\section{\VyZX}\label{sec:vyzx}

In designing \VyZX, we wanted to make it easy to write recursive or inductive functions over diagrams.
In a proof assistant like Coq, inductive structures allow for inductive proofs. 
Having the ability to use inductive proofs was a core goal for our definition as it would greatly simplify proofs.
For inspiration, we looked at diagrams for symmetric monoidal categories as described by Selinger~\cite{Selinger2010}.%
We reduced the basic requirements for our string diagrams to:
\begin{enumerate}
  \item The unit object, which is the empty diagram,
  \item The single wire,
  \item Morphisms, which take $n$ inputs to $m$ outputs,
  \item Braids, which swap two wires,
  \item Sequential composition, which composes two diagrams in sequence , and
  \item Tensor products, which arrange two diagrams in parallel.
\end{enumerate}

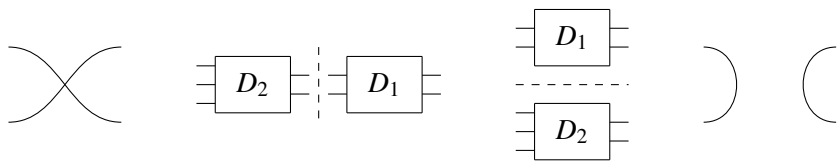
\begin{figure}[hb]
  \centering
  \input{swap-string.tikz}
  \qquad
  \input{compose-string.tikz}
  \qquad
  \input{parallel-string.tikz}
  \qquad
  \input{cap.tikz}
  \quad
  \input{cup.tikz}
  \caption{\centering From left to right, the braid, sequential composition, tensor product, cap, and cup for symmetric monoidal string diagrams.}\label{fig:string}
  \vspace{-5mm}
\end{figure}

These core objects give us our base language, consisting of a set of base morphisms, an empty diagram, a way to swap wires around, and the ability to compose diagrams in sequence and parallel.
When we wish to apply this to the ZX calculus, we must decide what our language's \emph{signature} will be. 
This signature constitutes the morphisms of our string diagram.
For the ZX calculus, a simple signature could include just the Z and X spiders.
We also include caps and cups in our signature to make diagrams easier to write, which are standard morphisms for string diagrams.
The building blocks for creating string diagrams inductively are given in \Cref{fig:string}, which we expand upon to build different representations for ZX diagrams.

\subsection{Block Representation ZX Diagrams}\label{sec:blk}

Our first goal with \VyZX is to create diagrams that can be used for proof in Coq.
We accomplish this by giving an inductive definition for ZX diagrams.
We refer to this representation as \textit{block representation ZX diagrams} in reference to how they may be stacked together and how stacks can line up with one another.
Each ZX diagram holds information about how many inputs and outputs it has, allowing us to define composition in a way that matches the outputs and inputs of two diagrams through a \texttt{Compose} constructor.
We also have a \texttt{Stack} operation that places one diagram on top of another.
Our base constructors are the \texttt{Z\_Spider}, \texttt{X\_Spider}, \texttt{Cap}, \texttt{Cup}, \texttt{Swap}, and \texttt{Empty} diagrams.
The type of a ZX diagram is given by \(\texttt{ZX} : \N \to \N \to \texttt{Type}\)
and the constructors are given by~\Cref{fig:blockconstructors}
\begin{figure}[t]
    \centering
    \begin{align*}
        \inferrule
        {\oftype{in out}{\(\N\)} \\ \oftype{\(\alpha\)}{\(\R\)}}
        {\oftype{\texttt{Z_Spider} in out \(\alpha\)}{\ZX{in}{out}}}
        \hspace{1em}&\hspace{1em}
        \inferrule
        {\oftype{in out}{\(\N\)} \\ \oftype{\(\alpha\)}{\(\R\)}}
        {\oftype{\texttt{X_Spider} in out \(\alpha\)}{\ZX{in}{out}}}
        \\
        \inferrule{ }{\oftype{\texttt{Cap}}{\ZX{0}{2}}}
        \hspace{2em}
        \inferrule{ }{\oftype{\texttt{Cup}}{\ZX{2}{0}}}
        \hspace{1em}&\hspace{1em}
        \inferrule{ }{\oftype{\texttt{Swap}}{\ZX{2}{2}}}
        \hspace{2em}
        \inferrule{ }{\oftype{\texttt{Empty}}{\ZX{0}{0}}}
        \\
        \inferrule
        {\oftype{zx1}{\ZX{in}{mid}} \\ \oftype{zx2}{\ZX{mid}{out}}}
        {\oftype{\texttt{Compose} zx1 zx2}{\ZX{in}{out}}}
        \hspace{1em}&\hspace{1em}
        \inferrule
        {\oftype{zx1}{\ZX{in1}{out1}} \\ \oftype{zx2}{\ZX{in2}{out2}}}
        {\oftype{\texttt{Stack} zx1 zx2}{\ZX{(in1 + in2)}{(out1 + out2)}}}
    \end{align*}
    \vspace*{-5mm}
    \caption{The inductive constructors for block representation ZX diagrams}\label{fig:blockconstructors}
    \vspace*{-5mm}
\end{figure}
These eight constructors allow us to write simple recursive functions and inductive proofs over ZX diagrams while allowing us to describe arbitrary diagrams.
Graphically these constructors correspond to the diagrams seen in~\Cref{fig:string} with the addition of the Z and X spiders shown in \Cref{fig:eqnXZ}.

We can view the same ZX diagrams that we describe using block representation not based on their building blocks but instead on nodes' adjacency. 
We refer to this view as the \textit{graph representation}.
In the following, we will see how graph representation can be useful for optimization and for proving certain rules.
In \Cref{sec:graph} we will discuss how we can convert block representation to graph representation.

\subsection{Semantics of diagrams} 
Given our inductive definition for ZX diagrams, we can write a simple function (\Cref{alg:sem}) for computing the semantics of a given ZX diagram.
We use the matrix definition given in \QLib~\cite{QuantumLib} to compute the semantics.
In \QLib a matrix is simply a function with type $\N \to \N \to \C$ that takes in a row and column index and returns the associated complex number.
First, we define the Z and X spider semantics, as stated in \Cref{sec:zx}, letting $\times$ be matrix multiplication\footnote{We chose to align our notation with \QLib rather than mathematical convention}, $I_{m\times n}$ be the $m$ by $n$ identity matrix, $H$ be the Hadamard matrix, $\otimes$ the Kronecker product, and $H^{\otimes n}$ be the $n$th power of $H$ with the Kronecker product.
With spider semantics complete, we define our other base constructors, stacks and composes using the Kronecker and matrix products.

\begin{algorithm}[h]
  \caption{ZX Diagram Semantics}\label{alg:sem}
  \scriptsize
  \begin{algorithmic}
    \Function{Z_Spider_semantics}{in, out, $\alpha$}
        \State \Return \(\begin{bmatrix}
                            1 & 0 & \cdots & 0 & 0\\
                            0 & 0 & \cdots & 0 & 0\\
                            \vdots & & \ddots & & \vdots\\
                            0 & 0 & \cdots & 0 & 0\\
                            0 & 0 & \cdots & 0 & e^{i\alpha}\\
                        \end{bmatrix}\)
                        \Comment Equivalent to \(\underbrace{\ket{0}\cdots\ket{0}}_{out}  \underbrace{\bra{0}\cdots\bra{0}}_{in} + e^{i\alpha} \underbrace{\ket{1}\cdots\ket{1}}_{out} \underbrace{\bra{1}\cdots\bra{1}}_{in}\)
    \EndFunction

  \Function{X_Spider_semantics}{in, out, $\alpha$}
    \State \Return \(H^{\otimes\text{out}}\) \(\times\) \Call{Z_Spider_semantics}{in, out, $\alpha$} \(\times\) \(H^{\otimes\text{in}}\)
    \Comment Equivalent to \(\underbrace{\ket{+}\cdots\ket{+}}_{out}  \underbrace{\bra{+}\cdots\bra{+}}_{in} + e^{i\alpha} \underbrace{\ket{-}\cdots\ket{-}}_{out} \underbrace{\bra{-}\cdots\bra{-}}_{in}\)
  \EndFunction
    \Function{ZX_semantics}{zx : ZX in out} : \(\C^{\text{in} \times \text{out}}\)
        \Switch{zx}
        \Case{Empty} 
          \State \Return $I_{1 \times 1}$ \EndCase
        \Case{Swap} 
          \State \Return \(\begin{bmatrix}1,0,0,0\\0,0,1,0\\0,1,0,0\\0,0,0,1\end{bmatrix}\) \EndCase
        \Case{Cap} 
          \State \Return \(\begin{bmatrix}1,0,0,1\end{bmatrix}^T\) \EndCase
        \Case{Cup} 
          \State \Return \(\begin{bmatrix}1,0,0,1\end{bmatrix}\) \EndCase
        \Case{Z_Spider in out $\alpha$} 
          \State \Return \Call{Z_Spider_semantics}{in, out, $\alpha$} \EndCase
        \Case{X_Spider in, out $\alpha$} 
          \State \Return \Call{X_Spider_semantics}{in, out, $\alpha$} \EndCase
        \Case{Stack zx1 zx2} 
          \State \Return \Call{ZX_semantics}{zx1} $\otimes$ \Call{ZX_semantics}{zx2} \EndCase
        \Case{Compose zx1 zx2} 
          \State \Return \Call{ZX_semantics}{zx2} $\times$ \Call{ZX_semantics}{zx1} \EndCase
        \EndSwitch
    \EndFunction
  \end{algorithmic}
\end{algorithm}

\subsection{Proportionality of diagrams}

Trivially two syntactically equal diagrams are to be considered equal.
For making useful statements, however, we require a notion of semantic equivalence.
Intuitively, one might define that as equivalence of matrices produced by \texttt{ZX_semantics} (as shown in \Cref{alg:sem}).
In the ZX calculus, though, we only care about equivalence up to multiplication by a constant factor, as rules will introduce constant factors and we are able to rebuild any constant factor if necessary using ZX constructions~\cite{vandewetering2020zxcalculus}.

Within \VyZX we define a relation called \textit{proportional} and give it the notation $\propto$.
We say 
$zx_1 \propto zx_2$ if there is a non-zero complex number $c$ such that 
\texttt{ZX_semantics} $zx_1$ = $c * {}$ \texttt{ZX_semantics} $zx_2$.
We prove that $\propto$ is an equivalence relation as we might expect.
We then prove that our composition operators respect proportionality: That is, if $zx_1 \propto zx_1'$ and $zx_2 \propto zx_2'$ then $\texttt{Compose}~zx_1~zx_2 \propto \texttt{Compose}~zx_1'~zx_2'$ and $\texttt{Stack}~zx_1~zx_2 \propto \texttt{Stack}~zx_1'~zx_2'$. We add this fact to Coq as a \emph{parametric morphism}, allowing us to rewrite using our equivalences even inside a broader diagram.
With proportionality defined, we proceed to verify different rewrite rules within the ZX calculus by proving their diagrams to be proportional.

\pagebreak
\subsection{Proving the Correctness of the ZX-Calculus}\label{sec:rules}
We can now show the rules to manipulate and simplify \VyZX diagrams.
For readability, we show most rules in graph representation.

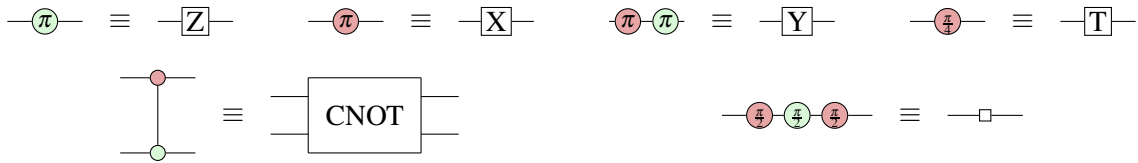
\begin{figure}[!h]
  \centering
  \input{gates.tikz}
  \caption{\centering Quantum gates represented in the ZX calculus. Note that the $H$ node omits the its label and is common to ZX diagrams as syntactic sugar for the rotations above.}\label{fig:gates}
  \vspace*{-5mm}
\end{figure}

\paragraph*{Common gates}
We translated common gates from the circuit model to the ZX-calculus and proved their semantic correctness, shown in \Cref{fig:gates}.

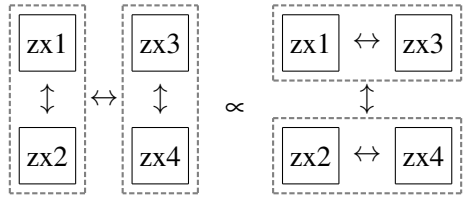
\begin{figure}[h]
  \centering
  \input{stack-compose-comm.tikz}
  \caption{\centering The distribution of stack (\(\updownarrow\)) and compose (\(\leftrightarrow\)).}\label{fig:stack-compose-comm}
\end{figure}

\paragraph*{Stack \& Compose distribute}

Sequential composition and stacking distribute as long as the individual diagrams have compatible dimensions by the rules stated in \Cref{sec:blk}.
\Cref{fig:stack-compose-comm} shows this property visually.
This fact is central to proving statements in block representation as it enables the diagram's structure to be changed while keeping the components the same.

\begin{wrapfigure}{r}{.45\textwidth}

    \begin{subfigure}{.45\textwidth}
        \centering
        \input{bihadamard.tikz}
        \caption{\centering Swapping a spider's color using a bi-Hadamard construction}
        \label{fig:bihadamard}
    \end{subfigure}
    
    \begin{subfigure}{.45\textwidth}
        \centering
        \input{bihadamard-diag.tikz}
        \caption{\centering Swapping colors of a diagram using a bi-Hadamard construction}
        \label{fig:bihadamard-diag}
    \end{subfigure}

    \caption{Color swapping ZX diagrams}

\end{wrapfigure}
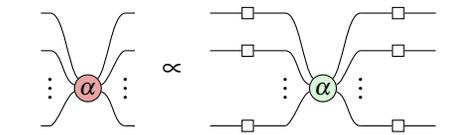
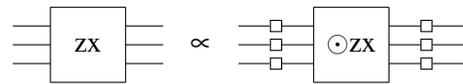

\paragraph*{Bi-Hadamard color changing}

We define a \emph{color-swapped ZX diagram} as a ZX diagram with the same structure but with all Z spiders being replaced by X spiders (while keeping the angle) and vice versa; all other constructions such as Cap, Cup, and Swap do not change~\cite{coecke-kissinger-2017-picturing-q-proc}.
Henceforth, we denote the color-swapped version of a ZX diagram \(\text{zx}\) as \(\odot \text{zx}\).
For a given spider, one can swap the spider's color while keeping the angle by composing a stack of Hadamards to the in and outputs.
This ``bi-Hadamard'' construction is shown in \Cref{fig:bihadamard}.
Further, we see this holds for all other non-compositional ZX diagram components (SWAPs, Caps, and Cups) since they do not have color and do not cause roration.%
We go on to show that, in fact, the bi-Hadamard rule is true for all color swapped ZX diagrams, as shown in \Cref{fig:bihadamard-diag}.

\paragraph*{Color swapping}

Using the previous fact about bi-Hadamard constructions, we prove that if a rule can be applied to a ZX diagram \(\text{zx}_1\) transforming it into \(\text{zx}_2\), then it can be applied to the color swapped diagram of \(\text{zx}_1\) transforming it into the color swapped diagram of \(\text{zx}_2\).
With this fact in mind, we only show one color configuration for any rule, understanding that it applies to the color-swapped version.
In practice, this allows us to prove any rule only for one color configuration and get the color-swapped lemma for free.
Since many proofs are computationally expensive, this greatly speeds up verification.

\begin{wrapfigure}[4]{r}{.3\textwidth}
    \centering
    \input{bi-alg.tikz}
\end{wrapfigure}

\paragraph*{Bi-algebra rule}
The bialgebra rule, while not intuitive, is crucial for many ZX proofs as it allows for the rearranging of edges between nodes
~\cite{vandewetering2020zxcalculus}.
Unlike the rules we have so far, it requires proof at matrix level and hence has a computationally expensive proof.
Though, once proven, it can be repeatedly applied, and given it rearranges edges can be used to prove many future facts.

\paragraph*{Hopf rule}

\begin{wrapfigure}[3]{r}{.25\textwidth}
    \centering
    \input{hopf.tikz}
\end{wrapfigure}

The Hopf rule, like the bialgebra rule, deals with the interaction between Z and X spiders.
It says that two edges between an X and a Z spider can be removed.
Intuitively, this tells us that no matter how much ``information'' we know about the X basis, given by our input, we do not get any information about the orthogonal Z basis~\cite{vandewetering2020zxcalculus}.
In practice, the Hopf rule allows us to disconnect specific nodes in the diagram instead of changing their connection.
As with the bi-algebra rule, this rule is proven directly on matrices and requires computation, though since the intermediate matrices are smaller this proof is less computationally expensive.

\paragraph*{Bi-\( \pi \) rule}

\begin{wrapfigure}[5]{r}{.3\textwidth}
    \centering
    \input{bi-pi.tikz}
\end{wrapfigure}

Any X spider is equal to itself, surrounded by Z rotations by \(\pi\).
Intuitively this rule is true due to the orthogonal nature of the X and Z basis and the fact that we are performing in total a full rotation.
A corollary of this rule is the \( \pi \)-copy rule~\cite{vandewetering2020zxcalculus} as shown and derived in the figure on the right.
The \( \pi \)-copy rule is not yet implemented, given it semantically builds on adjacency rather than a block like construction, since any input or output could be the one with the \(\pi\) spider on it. 
Generally speaking, rules that modify an arbitrary subset of the inputs and outputs are hard to represent in block representation and are left to be proven in for graph representation. %
Since the \( \pi \)-copy rule is what is usually presented in the literature, the bi-\( \pi \) is an interesting case study of a rule tailored to block representation that is equivalent to a more traditional rule.

\begin{figure}[ht]
    \centering
    \input{pi-copy.tikz}
    \caption{\centering The \( \pi \)-copy rule derivation using the bi-\( \pi \) rule}\label{fig:pi-copy}
    \vspace*{-5mm}
\end{figure}

\paragraph*{Spider fusion/splitting}

\begin{wrapfigure}[15]{r}{.45\textwidth}
    \begin{subfigure}{.45\textwidth}
    \centering
    \input{spider-fusion.tikz}
    \caption{\centering The spider fusion rule: Two connected nodes fused into a new node with added angles.}\label{fig:spider-fusion}
        \end{subfigure}
    
    \par
    
    \begin{subfigure}{.45\textwidth}
      \centering
      \input{spider-fusion-1-1.tikz}
      \caption{\centering A restricted version of spider fusion currently supported by \VyZX{}}\label{fig:spider-fusion-1-1}
    \end{subfigure}
    \caption{\centering Spider fusion rule}
\end{wrapfigure}
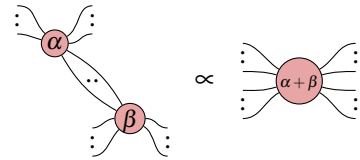
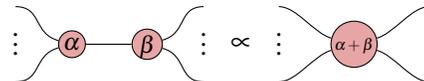 

One of the most important rules is that spiders connected by an arbitrary (non-zero) number of edges can be fused into a single node with the angles added~\cite{vandewetering2020zxcalculus}.
This rule is shown in \Cref{fig:spider-fusion}. 
Further, the reverse is true: any spider can be split such that the two new spiders add to the original angle.
A corollary of this is that we can split phase-gadgets (as discussed in \Cref{sec:zx-opt}) off nodes.

Since this rule fundamentally works based on adjacency rather than block representation construction, we have not fully implemented it at the time of writing.
We plan to follow with the general fusion rule, using the graph representation described in \Cref{sec:graph}.
We do, however, have a restricted version proven where two nodes are just connected to each other by a single wire, which will form the basis of the general spider fusion proof.
\Cref{fig:spider-fusion-1-1} shows this restricted version.
We implemented this by using the bra-ket semantics of spiders, as shown in \Cref{fig:eqnXZ}, instead of the direct matrix semantics shown in \Cref{alg:sem}.
This allows us to use \QLib's algebraic rewrites of complex matrices to combine the angles easily and shows why having both versions of our semantics  is valuable for proof.
Given the algebraic rewrite this rule is computationally very efficient.

\subsection{SQIR Ingestion}\label{sec:ingest}

We want \VyZX to be able to ingest arbitrary circuits.
To achieve this, \VyZX reads in circuits written in \SQIR, an intermediate representation for the \VOQC compiler embedded in Coq.
By choosing \SQIR we maintain interoperability with another verified compiler.
\SQIR represents circuits with \(q\) qubits as compositions of arbitrary \(x,y,z\) rotations acting on qubit \(n\) and CNOTs between arbitrary qubits \(n,m\), so this transformation and its verification is not trivial.

\begin{figure}[!h]

    \begin{subfigure}{.5\textwidth}
        \begin{subfigure}{\textwidth}
            \centering
            \input{shift-qubits-small.tikz}
            \caption{\centering Shifting qubits, the base construction for arbitrary swaps.}\label{fig:zx-shift}
        \end{subfigure}
        \begin{subfigure}{\textwidth}
            \centering
            ~\\[15pt]
            \input{ingest-rotation.tikz}
            \caption{\centering Constructing a rotation \(x,y,z\) as a ZX diagram using the Y gate from \Cref{fig:gates}}\label{fig:ingest-rotation}
        \end{subfigure}
    \end{subfigure}%
    \begin{subfigure}{.5\textwidth}
        \centering
        \input{ingest-cnot.tikz}
        \caption{\centering Constructing CNOT gate on qubits \(n\) and \(c\) for a circuit with \(q\) qubits (WLOG \(n < c\))}\label{fig:ingest-cnot}
    \end{subfigure}
    
    \caption{\centering Construction of SQIR operations in our ZX representation}
    \vspace*{-4mm}
\end{figure}
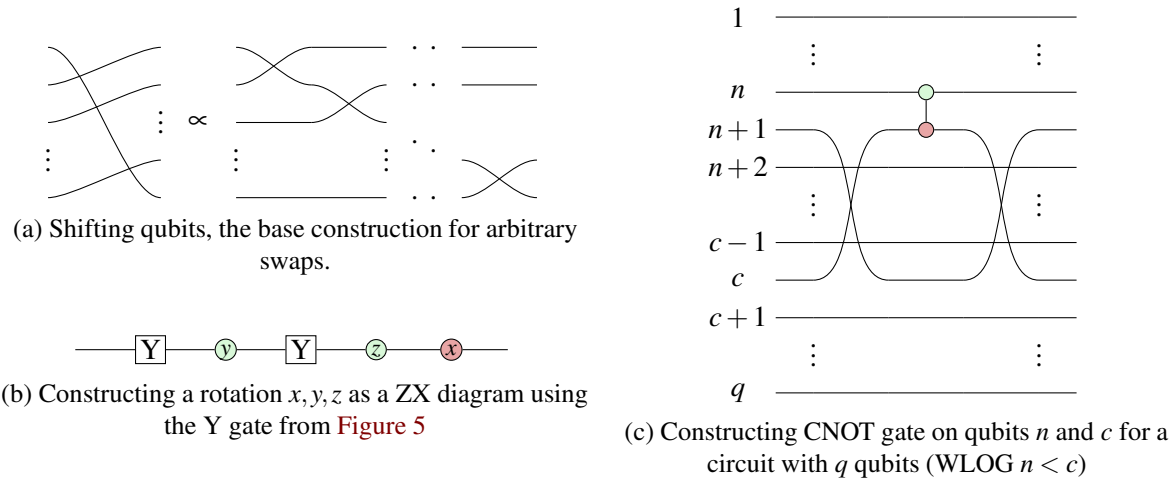

Circuit ingestion works by using arbitrary swaps, meaning instead of only having swap gates that swap two adjacent qubits, we have swap gates able to swap arbitrary two qubits\footnote{For the implementation we chose to have an IR with first-class arbitrary swaps that will then be translated into our base block representation IR (preserving semantics) using the following constructions.}. %
To construct an arbitrary swap, we first build a construction to shift qubit \(1\) to \(n\), whereby all other qubits are shifted up.
We also build the inverse (i.e., shifting qubit \(n\) to \(1\)).
\Cref{fig:zx-shift} shows this construction.
With this we can easily construct our arbitrary SWAP gate (shift \(1\) to \(n\), then \(n-1\) to \(1\)).
The discussion below will be in block representation that has defined swaps (see sec \Cref{sec:blk}) of two adjacent qubits as the composition of \(3\) CNOTs.

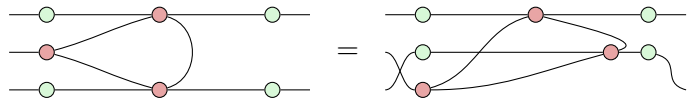
\begin{wrapfigure}[9]{r}{0.6\textwidth}
    \centering
    \input{deformation.tikz}
    \vspace*{-5mm}
    \caption{\centering Two equal ZX diagrams, where the right diagram is deformed. Note that all connections, inputs/outputs, and qubit order are maintained.}\label{fig:deform}
\end{wrapfigure}

Using arbitrary swaps and shifts, we can now interpret any wire crossing.
Hence, in block representation we can construct a CNOT acting on two qubits \(n,c\) by swapping one qubit next to the other qubit, applying the CNOT, and swapping back, as shown in \Cref{fig:ingest-cnot}.
To then convert such a circuit, \texttt{VyZX} converts rotations \(x, y, z\) into the construction shown in \Cref{fig:ingest-rotation}.
Composition of \texttt{SQIR} terms is also represented by composition in our block representation IR.

%% file: figures/swap-string.tikz
\begin{tikzpicture}
	\begin{pgfonlayer}{nodelayer}
		\node [style=none] (0) at (-1.5, 1) {};
		\node [style=none] (1) at (-1.5, -1) {};
		\node [style=none] (2) at (1.5, 1) {};
		\node [style=none] (3) at (1.5, -1) {};
	\end{pgfonlayer}
	\begin{pgfonlayer}{edgelayer}
		\draw [in=-180, out=0] (0.center) to (3.center);
		\draw [in=0, out=180] (2.center) to (1.center);
	\end{pgfonlayer}
\end{tikzpicture}

%% file: figures/compose-string.tikz
\begin{tikzpicture}
	\begin{pgfonlayer}{nodelayer}
		\node [style=none] (4) at (0, 1) {};
		\node [style=none] (5) at (0, -1) {};
		\node [style=none] (7) at (2.75, -0.75) {};
		\node [style=none] (8) at (3.25, -0.25) {};
		\node [style=none] (9) at (3.25, 0.25) {};
		\node [style=none] (10) at (2.75, 0.75) {};
		\node [style=none] (11) at (-0.75, 0.75) {};
		\node [style=none] (12) at (-0.25, 0.25) {};
		\node [style=none] (13) at (-0.25, -0.25) {};
		\node [style=none] (14) at (-0.75, -0.75) {};
		\node [style=none] (15) at (-3.25, 0.5) {};
		\node [style=none] (16) at (-3.25, 0) {};
		\node [style=none] (17) at (-3.25, -0.5) {};
		\node [style=none] (18) at (-2.75, 0.5) {};
		\node [style=none] (19) at (-2.75, 0) {};
		\node [style=none] (20) at (-2.75, -0.5) {};
		\node [style=none] (21) at (0.75, -0.75) {};
		\node [style=none] (22) at (0.25, -0.25) {};
		\node [style=none] (23) at (0.25, 0.25) {};
		\node [style=none] (24) at (0.75, 0.75) {};
		\node [style=none] (25) at (0.75, 0.25) {};
		\node [style=none] (26) at (0.75, -0.25) {};
		\node [style=none] (27) at (2.75, -0.25) {};
		\node [style=none] (28) at (2.75, 0.25) {};
		\node [style=none] (29) at (-2.75, 0.75) {};
		\node [style=none] (30) at (-2.75, -0.75) {};
		\node [style=none] (31) at (-0.75, 0.25) {};
		\node [style=none] (32) at (-0.75, -0.25) {};
		\node [style=none] (33) at (1.75, 0) {$D_1$};
		\node [style=none] (34) at (-1.75, 0) {$D_2$};
	\end{pgfonlayer}
	\begin{pgfonlayer}{edgelayer}
		\draw [style=dashed] (4.center) to (5.center);
		\draw (24.center) to (10.center);
		\draw (10.center) to (7.center);
		\draw (7.center) to (21.center);
		\draw (21.center) to (24.center);
		\draw (23.center) to (25.center);
		\draw (22.center) to (26.center);
		\draw (28.center) to (9.center);
		\draw (27.center) to (8.center);
		\draw (29.center) to (11.center);
		\draw (11.center) to (14.center);
		\draw (14.center) to (30.center);
		\draw (30.center) to (29.center);
		\draw (15.center) to (18.center);
		\draw (16.center) to (19.center);
		\draw (17.center) to (20.center);
		\draw (31.center) to (12.center);
		\draw (32.center) to (13.center);
	\end{pgfonlayer}
\end{tikzpicture}

%% file: figures/parallel-string.tikz
\begin{tikzpicture}
	\begin{pgfonlayer}{nodelayer}
		\node [style=none] (4) at (-1.5, 0) {};
		\node [style=none] (5) at (1.5, 0) {};
		\node [style=none] (7) at (1, 0.5) {};
		\node [style=none] (8) at (1.5, 1) {};
		\node [style=none] (9) at (1.5, 1.5) {};
		\node [style=none] (10) at (1, 2) {};
		\node [style=none] (11) at (1, -0.5) {};
		\node [style=none] (12) at (1.5, -1) {};
		\node [style=none] (13) at (1.5, -1.5) {};
		\node [style=none] (14) at (1, -2) {};
		\node [style=none] (15) at (-1.5, -0.75) {};
		\node [style=none] (16) at (-1.5, -1.25) {};
		\node [style=none] (17) at (-1.5, -1.75) {};
		\node [style=none] (18) at (-1, -0.75) {};
		\node [style=none] (19) at (-1, -1.25) {};
		\node [style=none] (20) at (-1, -1.75) {};
		\node [style=none] (21) at (-1, 0.5) {};
		\node [style=none] (22) at (-1.5, 1) {};
		\node [style=none] (23) at (-1.5, 1.5) {};
		\node [style=none] (24) at (-1, 2) {};
		\node [style=none] (25) at (-1, 1.5) {};
		\node [style=none] (26) at (-1, 1) {};
		\node [style=none] (27) at (1, 1) {};
		\node [style=none] (28) at (1, 1.5) {};
		\node [style=none] (29) at (-1, -0.5) {};
		\node [style=none] (30) at (-1, -2) {};
		\node [style=none] (31) at (1, -1) {};
		\node [style=none] (32) at (1, -1.5) {};
		\node [style=none] (33) at (0, 1.25) {$D_1$};
		\node [style=none] (34) at (0, -1.25) {$D_2$};
	\end{pgfonlayer}
	\begin{pgfonlayer}{edgelayer}
		\draw [style=dashed] (4.center) to (5.center);
		\draw (24.center) to (10.center);
		\draw (10.center) to (7.center);
		\draw (7.center) to (21.center);
		\draw (21.center) to (24.center);
		\draw (23.center) to (25.center);
		\draw (22.center) to (26.center);
		\draw (28.center) to (9.center);
		\draw (27.center) to (8.center);
		\draw (29.center) to (11.center);
		\draw (11.center) to (14.center);
		\draw (14.center) to (30.center);
		\draw (30.center) to (29.center);
		\draw (15.center) to (18.center);
		\draw (16.center) to (19.center);
		\draw (17.center) to (20.center);
		\draw (31.center) to (12.center);
		\draw (32.center) to (13.center);
	\end{pgfonlayer}
\end{tikzpicture}

%% file: figures/cap.tikz
\begin{tikzpicture}
	\begin{pgfonlayer}{nodelayer}
		\node [style=none] (0) at (-0.5, 1) {};
		\node [style=none] (1) at (-0.5, -1) {};
	\end{pgfonlayer}
	\begin{pgfonlayer}{edgelayer}
		\draw [bend left=90, looseness=1.50] (0.center) to (1.center);
	\end{pgfonlayer}
\end{tikzpicture}

%% file: figures/cup.tikz
\begin{tikzpicture}
	\begin{pgfonlayer}{nodelayer}
		\node [style=none] (2) at (0.5, 1) {};
		\node [style=none] (3) at (0.5, -1) {};
	\end{pgfonlayer}
	\begin{pgfonlayer}{edgelayer}
		\draw [bend right=90, looseness=1.50] (2.center) to (3.center);
	\end{pgfonlayer}
\end{tikzpicture}

%% file: figures/gates.tikz
\begin{tikzpicture}
	\begin{pgfonlayer}{nodelayer}
		\node [style=none] (2) at (29, 0) {};
		\node [style=none] (3) at (27, 0) {};
		\node [style=none] (4) at (3, 0) {};
		\node [style=none] (5) at (5, 0) {};
		\node [style=white dot] (18) at (28, 0) {\tiny{}$\frac{\pi}{4}$};
		\node [style=gray dot] (19) at (4, 0) {$\pi$};
		\node [style=none] (29) at (11, 0) {};
		\node [style=none] (30) at (13, 0) {};
		\node [style=none] (31) at (19, 0) {};
		\node [style=none] (32) at (21, 0) {};
		\node [style=white dot] (33) at (12, 0) {$\pi$};
		\node [style=white dot] (36) at (19.5, 0) {$\pi$};
		\node [style=gray dot] (37) at (20.5, 0) {$\pi$};
		\node [style=none] (38) at (8, -1.5) {};
		\node [style=none] (39) at (8, -3.5) {};
		\node [style=white dot] (40) at (7, -1.5) {};
		\node [style=gray dot] (41) at (7, -3.5) {};
		\node [style=none] (42) at (6, -1.5) {};
		\node [style=none] (43) at (6, -3.5) {};
		\node [style=none] (45) at (22, -2.5) {};
		\node [style=white dot] (46) at (23, -2.5) {\tiny{}$\frac{\pi}{2}$};
		\node [style=white dot] (47) at (25, -2.5) {\tiny{}$\frac{\pi}{2}$};
		\node [style=gray dot] (48) at (24, -2.5) {\tiny{}$\frac{\pi}{2}$};
		\node [style=none] (49) at (26, -2.5) {};
		\node [style=none] (51) at (28, -2.5) {};
		\node [style=none] (52) at (30, -2.5) {};
		\node [style=hadamard] (53) at (29, -2.5) {};
		\node [style=none] (54) at (9, -2.5) {$\equiv$};
		\node [style=none] (55) at (22, 0) {$\equiv$};
		\node [style=none] (56) at (23, 0) {};
		\node [style=none] (57) at (25, 0) {};
		\node [style=hadamard] (58) at (24, 0) {Y};
		\node [style=none] (59) at (27, -2.5) {$\equiv$};
		\node [style=none] (60) at (31, 0) {};
		\node [style=none] (61) at (33, 0) {};
		\node [style=hadamard] (62) at (32, 0) {T};
		\node [style=none] (63) at (30, 0) {$\equiv$};
		\node [style=none] (64) at (14, 0) {$\equiv$};
		\node [style=none] (65) at (15, 0) {};
		\node [style=none] (66) at (17, 0) {};
		\node [style=hadamard] (67) at (16, 0) {X};
		\node [style=none] (68) at (6, 0) {$\equiv$};
		\node [style=none] (69) at (7, 0) {};
		\node [style=none] (70) at (9, 0) {};
		\node [style=hadamard] (71) at (8, 0) {Z};
		\node [style=none] (72) at (11, -1.5) {};
		\node [style=none] (73) at (14, -1.5) {};
		\node [style=none] (74) at (14, -3.5) {};
		\node [style=none] (75) at (11, -3.5) {};
		\node [style=none] (76) at (11, -2) {};
		\node [style=none] (77) at (14, -2) {};
		\node [style=none] (78) at (15, -2) {};
		\node [style=none] (79) at (15, -3) {};
		\node [style=none] (80) at (14, -3) {};
		\node [style=none] (81) at (11, -3) {};
		\node [style=none] (82) at (10, -3) {};
		\node [style=none] (83) at (10, -2) {};
		\node [style=none] (84) at (12.5, -2.5) {CNOT};
	\end{pgfonlayer}
	\begin{pgfonlayer}{edgelayer}
		\draw (32.center) to (37);
		\draw (37) to (36);
		\draw (36) to (31.center);
		\draw (29.center) to (33);
		\draw (33) to (30.center);
		\draw (4.center) to (19);
		\draw (19) to (5.center);
		\draw (2.center) to (18);
		\draw (18) to (3.center);
		\draw (42.center) to (40);
		\draw (40) to (38.center);
		\draw (43.center) to (41);
		\draw (41) to (39.center);
		\draw (40) to (41);
		\draw (51.center) to (53);
		\draw (53) to (52.center);
		\draw (49.center) to (47);
		\draw (47) to (48);
		\draw [in=0, out=-180] (48) to (46);
		\draw (46) to (45.center);
		\draw (56.center) to (58);
		\draw (58) to (57.center);
		\draw (60.center) to (62);
		\draw (62) to (61.center);
		\draw (65.center) to (67);
		\draw (67) to (66.center);
		\draw (69.center) to (71);
		\draw (71) to (70.center);
		\draw (72.center) to (73.center);
		\draw (73.center) to (74.center);
		\draw (74.center) to (75.center);
		\draw (75.center) to (72.center);
		\draw (83.center) to (76.center);
		\draw (81.center) to (82.center);
		\draw (78.center) to (77.center);
		\draw (79.center) to (80.center);
	\end{pgfonlayer}
\end{tikzpicture}

%% file: figures/stack-compose-comm.tikz
\begin{tikzpicture}
	\begin{pgfonlayer}{nodelayer}
		\node [style=none] (0) at (-5.75, 2.5) {};
		\node [style=none] (1) at (-5.75, 1) {};
		\node [style=none] (2) at (-4.25, 1) {};
		\node [style=none] (3) at (-4.25, 2.5) {};
		\node [style=none] (4) at (-5, 0.25) {$\updownarrow$};
		\node [style=none] (5) at (-5, 1.75) {zx1};
		\node [style=none] (23) at (-3.5, 0.25) {$\leftrightarrow$};
		\node [style=none] (24) at (-5.75, -0.5) {};
		\node [style=none] (25) at (-5.75, -2) {};
		\node [style=none] (26) at (-4.25, -2) {};
		\node [style=none] (27) at (-4.25, -0.5) {};
		\node [style=none] (28) at (-5, -1.25) {zx2};
		\node [style=none] (29) at (-2.75, 2.5) {};
		\node [style=none] (30) at (-2.75, 1) {};
		\node [style=none] (31) at (-1.25, 1) {};
		\node [style=none] (32) at (-1.25, 2.5) {};
		\node [style=none] (33) at (-2, 0.25) {$\updownarrow$};
		\node [style=none] (34) at (-2, 1.75) {zx3};
		\node [style=none] (35) at (-2.75, -0.5) {};
		\node [style=none] (36) at (-2.75, -2) {};
		\node [style=none] (37) at (-1.25, -2) {};
		\node [style=none] (38) at (-1.25, -0.5) {};
		\node [style=none] (39) at (-2, -1.25) {zx4};
		\node [style=none] (40) at (-4, 2.75) {};
		\node [style=none] (41) at (-6, 2.75) {};
		\node [style=none] (42) at (-6, -2.25) {};
		\node [style=none] (43) at (-4, -2.25) {};
		\node [style=none] (44) at (-1, 2.75) {};
		\node [style=none] (45) at (-3, 2.75) {};
		\node [style=none] (46) at (-3, -2.25) {};
		\node [style=none] (47) at (-1, -2.25) {};
		\node [style=none] (48) at (1.25, 2.5) {};
		\node [style=none] (49) at (1.25, 1) {};
		\node [style=none] (50) at (2.75, 1) {};
		\node [style=none] (51) at (2.75, 2.5) {};
		\node [style=none] (52) at (1.25, -0.5) {};
		\node [style=none] (53) at (1.25, -2) {};
		\node [style=none] (54) at (2.75, -2) {};
		\node [style=none] (55) at (2.75, -0.5) {};
		\node [style=none] (56) at (4.25, 2.5) {};
		\node [style=none] (57) at (4.25, 1) {};
		\node [style=none] (58) at (5.75, 1) {};
		\node [style=none] (59) at (5.75, 2.5) {};
		\node [style=none] (60) at (4.25, -0.5) {};
		\node [style=none] (61) at (4.25, -2) {};
		\node [style=none] (62) at (5.75, -2) {};
		\node [style=none] (63) at (5.75, -0.5) {};
		\node [style=none] (64) at (2, 1.75) {zx1};
		\node [style=none] (65) at (2, -1.25) {zx2};
		\node [style=none] (66) at (5, 1.75) {zx3};
		\node [style=none] (67) at (5, -1.25) {zx4};
		\node [style=none] (68) at (3.5, 1.75) {$\leftrightarrow$};
		\node [style=none] (69) at (3.5, -1.25) {$\leftrightarrow$};
		\node [style=none] (70) at (3.5, 0.25) {$\updownarrow$};
		\node [style=none] (71) at (1, -0.25) {};
		\node [style=none] (72) at (1, -2.25) {};
		\node [style=none] (73) at (6, -2.25) {};
		\node [style=none] (74) at (6, -0.25) {};
		\node [style=none] (75) at (1, 2.75) {};
		\node [style=none] (76) at (1, 0.75) {};
		\node [style=none] (77) at (6, 0.75) {};
		\node [style=none] (78) at (6, 2.75) {};
		\node [style=none] (79) at (0, 0) {$\propto$};
	\end{pgfonlayer}
	\begin{pgfonlayer}{edgelayer}
		\draw [style=simple] (0.center) to (3.center);
		\draw [style=simple] (3.center) to (2.center);
		\draw [style=simple] (2.center) to (1.center);
		\draw [style=simple] (1.center) to (0.center);
		\draw [style=simple] (24.center) to (27.center);
		\draw [style=simple] (27.center) to (26.center);
		\draw [style=simple] (26.center) to (25.center);
		\draw [style=simple] (25.center) to (24.center);
		\draw [style=simple] (29.center) to (32.center);
		\draw [style=simple] (32.center) to (31.center);
		\draw [style=simple] (31.center) to (30.center);
		\draw [style=simple] (30.center) to (29.center);
		\draw [style=simple] (35.center) to (38.center);
		\draw [style=simple] (38.center) to (37.center);
		\draw [style=simple] (37.center) to (36.center);
		\draw [style=simple] (36.center) to (35.center);
		\draw [style=hadamard edge] (40.center) to (43.center);
		\draw [style=hadamard edge] (43.center) to (42.center);
		\draw [style=hadamard edge] (42.center) to (41.center);
		\draw [style=hadamard edge] (41.center) to (40.center);
		\draw [style=hadamard edge] (44.center) to (47.center);
		\draw [style=hadamard edge] (47.center) to (46.center);
		\draw [style=hadamard edge] (46.center) to (45.center);
		\draw [style=hadamard edge] (45.center) to (44.center);
		\draw [style=simple] (48.center) to (51.center);
		\draw [style=simple] (51.center) to (50.center);
		\draw [style=simple] (50.center) to (49.center);
		\draw [style=simple] (49.center) to (48.center);
		\draw [style=simple] (52.center) to (55.center);
		\draw [style=simple] (55.center) to (54.center);
		\draw [style=simple] (54.center) to (53.center);
		\draw [style=simple] (53.center) to (52.center);
		\draw [style=simple] (56.center) to (59.center);
		\draw [style=simple] (59.center) to (58.center);
		\draw [style=simple] (58.center) to (57.center);
		\draw [style=simple] (57.center) to (56.center);
		\draw [style=simple] (60.center) to (63.center);
		\draw [style=simple] (63.center) to (62.center);
		\draw [style=simple] (62.center) to (61.center);
		\draw [style=simple] (61.center) to (60.center);
		\draw [style=hadamard edge] (71.center) to (74.center);
		\draw [style=hadamard edge] (74.center) to (73.center);
		\draw [style=hadamard edge] (73.center) to (72.center);
		\draw [style=hadamard edge] (72.center) to (71.center);
		\draw [style=hadamard edge] (75.center) to (78.center);
		\draw [style=hadamard edge] (78.center) to (77.center);
		\draw [style=hadamard edge] (77.center) to (76.center);
		\draw [style=hadamard edge] (76.center) to (75.center);
	\end{pgfonlayer}
\end{tikzpicture}

%% file: figures/bihadamard.tikz
\begin{tikzpicture}
	\begin{pgfonlayer}{nodelayer}
		\node [style=none] (0) at (-1.25, 1) {};
		\node [style=white dot] (1) at (-0.25, 0) {$\alpha$};
		\node [style=none] (2) at (-1.25, 2) {};
		\node [style=none] (3) at (-1.25, -1) {};
		\node [style=none] (4) at (0.75, -1) {};
		\node [style=none] (5) at (0.75, 1) {};
		\node [style=none] (6) at (0.75, 2) {};
		\node [style=none] (7) at (0.75, 0.25) {.};
		\node [style=none] (8) at (0.75, 0) {.};
		\node [style=none] (9) at (0.75, -0.25) {.};
		\node [style=none] (10) at (-1.25, -0.25) {.};
		\node [style=none] (11) at (-1.25, 0) {.};
		\node [style=none] (12) at (-1.25, 0.25) {.};
		\node [style=none] (13) at (5, 1) {};
		\node [style=gray dot] (14) at (6, 0) {$\alpha$};
		\node [style=none] (15) at (5, 2) {};
		\node [style=none] (16) at (5, -1) {};
		\node [style=none] (17) at (7, -1) {};
		\node [style=none] (18) at (7, 1) {};
		\node [style=none] (19) at (7, 2) {};
		\node [style=none] (20) at (7, 0.25) {.};
		\node [style=none] (21) at (7, 0) {.};
		\node [style=none] (22) at (7, -0.25) {.};
		\node [style=none] (23) at (5, -0.25) {.};
		\node [style=none] (24) at (5, 0) {.};
		\node [style=none] (25) at (5, 0.25) {.};
		\node [style=hadamard] (26) at (4, 2) {};
		\node [style=hadamard] (27) at (4, 1) {};
		\node [style=hadamard] (28) at (4, -1) {};
		\node [style=hadamard] (29) at (8, -1) {};
		\node [style=hadamard] (30) at (8, 1) {};
		\node [style=hadamard] (31) at (8, 2) {};
		\node [style=none] (32) at (3, 2) {};
		\node [style=none] (33) at (3, 1) {};
		\node [style=none] (34) at (3, -1) {};
		\node [style=none] (35) at (2, 0.5) {$\propto$};
		\node [style=none] (36) at (9, 2) {};
		\node [style=none] (37) at (9, 1) {};
		\node [style=none] (38) at (9, -1) {};
		\node [style=none] (39) at (1, 2) {};
		\node [style=none] (40) at (1, 1) {};
		\node [style=none] (41) at (1, -1) {};
		\node [style=none] (42) at (-1.5, -1) {};
		\node [style=none] (43) at (-1.5, 1) {};
		\node [style=none] (44) at (-1.5, 2) {};
	\end{pgfonlayer}
	\begin{pgfonlayer}{edgelayer}
		\draw [in=135, out=0, looseness=0.50] (2.center) to (1);
		\draw [in=-180, out=45, looseness=0.50] (1) to (6.center);
		\draw [in=-180, out=15, looseness=0.50] (1) to (5.center);
		\draw [in=165, out=0, looseness=0.75] (0.center) to (1);
		\draw [in=15, out=-165, looseness=0.50] (1) to (3.center);
		\draw [in=-180, out=-15, looseness=0.75] (1) to (4.center);
		\draw [in=135, out=0, looseness=0.50] (15.center) to (14);
		\draw [in=-180, out=45, looseness=0.50] (14) to (19.center);
		\draw [in=-180, out=15, looseness=0.50] (14) to (18.center);
		\draw [in=165, out=0, looseness=0.75] (13.center) to (14);
		\draw [in=15, out=-165, looseness=0.50] (14) to (16.center);
		\draw [in=-180, out=-15, looseness=0.75] (14) to (17.center);
		\draw (19.center) to (31);
		\draw (31) to (36.center);
		\draw (37.center) to (30);
		\draw (30) to (18.center);
		\draw (17.center) to (29);
		\draw (29) to (38.center);
		\draw (16.center) to (28);
		\draw (33.center) to (27);
		\draw (27) to (13.center);
		\draw (15.center) to (26);
		\draw (26) to (32.center);
		\draw (34.center) to (28);
		\draw (44.center) to (2.center);
		\draw (43.center) to (0.center);
		\draw (42.center) to (3.center);
		\draw (41.center) to (4.center);
		\draw (40.center) to (5.center);
		\draw (39.center) to (6.center);
	\end{pgfonlayer}
\end{tikzpicture}

%% file: figures/bihadamard-diag.tikz
\begin{tikzpicture}
	\begin{pgfonlayer}{nodelayer}
		\node [style=none] (0) at (-2, 1) {};
		\node [style=none] (1) at (-2, -1) {};
		\node [style=none] (2) at (0, -1) {};
		\node [style=none] (3) at (0, 1) {};
		\node [style=none] (4) at (1, 0) {};
		\node [style=none] (5) at (1, 0.5) {};
		\node [style=none] (6) at (1, -0.5) {};
		\node [style=none] (7) at (-3, -0.5) {};
		\node [style=none] (8) at (-3, 0) {};
		\node [style=none] (9) at (-3, 0.5) {};
		\node [style=none] (10) at (0, 0) {};
		\node [style=none] (11) at (0, 0.5) {};
		\node [style=none] (12) at (0, -0.5) {};
		\node [style=none] (13) at (-2, 0) {};
		\node [style=none] (14) at (-2, 0.5) {};
		\node [style=none] (15) at (-2, -0.5) {};
		\node [style=none] (16) at (-1, 0) {};
		\node [style=none] (17) at (-1, 0) {zx};
		\node [style=none] (18) at (5, 1) {};
		\node [style=none] (19) at (5, -1) {};
		\node [style=none] (20) at (7, -1) {};
		\node [style=none] (21) at (7, 1) {};
		\node [style=hadamard] (22) at (8, 0) {};
		\node [style=hadamard] (23) at (8, 0.5) {};
		\node [style=hadamard] (24) at (8, -0.5) {};
		\node [style=hadamard] (25) at (4, -0.5) {};
		\node [style=hadamard] (26) at (4, 0) {};
		\node [style=hadamard] (27) at (4, 0.5) {};
		\node [style=none] (28) at (7, 0) {};
		\node [style=none] (29) at (7, 0.5) {};
		\node [style=none] (30) at (7, -0.5) {};
		\node [style=none] (31) at (5, 0) {};
		\node [style=none] (32) at (5, 0.5) {};
		\node [style=none] (33) at (5, -0.5) {};
		\node [style=none] (34) at (6, 0) {};
		\node [style=none] (35) at (6, 0) {$\odot$zx};
		\node [style=none] (36) at (9, 0) {};
		\node [style=none] (37) at (9, 0.5) {};
		\node [style=none] (38) at (9, -0.5) {};
		\node [style=none] (39) at (3, 0) {};
		\node [style=none] (40) at (3, 0.5) {};
		\node [style=none] (41) at (3, -0.5) {};
		\node [style=none] (42) at (2, 0) {$\propto$};
	\end{pgfonlayer}
	\begin{pgfonlayer}{edgelayer}
		\draw [style=simple] (0.center) to (3.center);
		\draw [style=simple] (3.center) to (2.center);
		\draw [style=simple] (2.center) to (1.center);
		\draw [style=simple] (1.center) to (0.center);
		\draw [style=simple] (11.center) to (5.center);
		\draw [style=simple] (10.center) to (4.center);
		\draw [style=simple] (6.center) to (12.center);
		\draw [style=simple] (15.center) to (7.center);
		\draw [style=simple] (8.center) to (13.center);
		\draw [style=simple] (14.center) to (9.center);
		\draw [style=simple] (18.center) to (21.center);
		\draw [style=simple] (21.center) to (20.center);
		\draw [style=simple] (20.center) to (19.center);
		\draw [style=simple] (19.center) to (18.center);
		\draw [style=simple] (29.center) to (23);
		\draw [style=simple] (28.center) to (22);
		\draw [style=simple] (24) to (30.center);
		\draw [style=simple] (33.center) to (25);
		\draw [style=simple] (26) to (31.center);
		\draw [style=simple] (32.center) to (27);
		\draw [style=simple] (27) to (40.center);
		\draw [style=simple] (39.center) to (26);
		\draw [style=simple] (41.center) to (25);
		\draw [style=simple] (24) to (38.center);
		\draw [style=simple] (22) to (36.center);
		\draw [style=simple] (23) to (37.center);
	\end{pgfonlayer}
\end{tikzpicture}

%% file: figures/bi-alg.tikz
\begin{tikzpicture}
	\begin{pgfonlayer}{nodelayer}
		\node [style=white dot] (0) at (-2, 0) {};
		\node [style=gray dot] (1) at (-3, 0) {};
		\node [style=none] (2) at (-1, 0.75) {};
		\node [style=none] (6) at (-1, -0.75) {};
		\node [style=none] (9) at (0, 0) {$\propto$};
		\node [style=none] (12) at (1, -0.75) {};
		\node [style=none] (13) at (1, 0.75) {};
		\node [style=white dot] (14) at (1.5, -0.75) {};
		\node [style=white dot] (15) at (1.5, 0.75) {};
		\node [style=gray dot] (16) at (3.5, 0.75) {};
		\node [style=gray dot] (17) at (3.5, -0.75) {};
		\node [style=none] (18) at (4, 0.75) {};
		\node [style=none] (19) at (4, -0.75) {};
		\node [style=none] (20) at (-4, 0.75) {};
		\node [style=none] (24) at (-4, -0.75) {};
	\end{pgfonlayer}
	\begin{pgfonlayer}{edgelayer}
		\draw (13.center) to (15);
		\draw (15) to (16);
		\draw [in=165, out=-15] (15) to (17);
		\draw [in=-165, out=15, looseness=1.25] (14) to (16);
		\draw (16) to (18.center);
		\draw (17) to (19.center);
		\draw (14) to (17);
		\draw (12.center) to (14);
		\draw [in=165, out=0] (20.center) to (1);
		\draw (1) to (0);
		\draw [in=-180, out=15] (0) to (2.center);
		\draw [in=-180, out=-15] (0) to (6.center);
		\draw [in=0, out=-180, looseness=0.75] (1) to (24.center);
	\end{pgfonlayer}
\end{tikzpicture}

%% file: figures/hopf.tikz
\begin{tikzpicture}
	\begin{pgfonlayer}{nodelayer}
		\node [style=white dot] (0) at (-3.25, 0) {};
		\node [style=gray dot] (1) at (-1.75, 0) {};
		\node [style=none] (2) at (-1, 0) {};
		\node [style=none] (3) at (-4, 0) {};
		\node [style=white dot] (4) at (1.75, 0) {};
		\node [style=gray dot] (5) at (3.25, 0) {};
		\node [style=none] (6) at (4, 0) {};
		\node [style=none] (7) at (1, 0) {};
		\node [style=none] (8) at (0, 0) {$\propto$};
	\end{pgfonlayer}
	\begin{pgfonlayer}{edgelayer}
		\draw (3.center) to (0);
		\draw [bend left=15, looseness=1.25] (0) to (1);
		\draw (1) to (2.center);
		\draw [in=-165, out=-15] (0) to (1);
		\draw (7.center) to (4);
		\draw (5) to (6.center);
	\end{pgfonlayer}
\end{tikzpicture}

%% file: figures/bi-pi.tikz
\begin{tikzpicture}
	\begin{pgfonlayer}{nodelayer}
		\node [style=none] (1) at (-1, 1) {};
		\node [style=white dot, minimum size=.66cm] (2) at (0, 0) {$-\alpha$};
		\node [style=none] (3) at (-1, 2) {};
		\node [style=none] (4) at (-1, -1) {};
		\node [style=none] (5) at (1, -1) {};
		\node [style=none] (6) at (1, 1) {};
		\node [style=none] (7) at (1, 2) {};
		\node [style=none] (8) at (1, 0.25) {.};
		\node [style=none] (9) at (1, 0) {.};
		\node [style=none] (10) at (1, -0.25) {.};
		\node [style=none] (11) at (-1, -0.25) {.};
		\node [style=none] (12) at (-1, 0) {.};
		\node [style=none] (13) at (-1, 0.25) {.};
		\node [style=gray dot] (14) at (1.5, 2) {$\pi$};
		\node [style=gray dot] (15) at (1.5, 1) {$\pi$};
		\node [style=gray dot] (16) at (1.5, -1) {$\pi$};
		\node [style=gray dot] (17) at (-1.5, -1) {$\pi$};
		\node [style=gray dot] (18) at (-1.5, 1) {$\pi$};
		\node [style=gray dot] (19) at (-1.5, 2) {$\pi$};
		\node [style=none] (20) at (2, 2) {};
		\node [style=none] (21) at (2, 1) {};
		\node [style=none] (22) at (2, -1) {};
		\node [style=none] (23) at (-2, 2) {};
		\node [style=none] (24) at (-2, 1) {};
		\node [style=none] (25) at (-2, -1) {};
		\node [style=none] (26) at (-6.25, 1) {};
		\node [style=white dot, minimum size=.66cm] (27) at (-5.25, 0) {$\alpha$};
		\node [style=none] (28) at (-6.25, 2) {};
		\node [style=none] (29) at (-6.25, -1) {};
		\node [style=none] (30) at (-4.25, -1) {};
		\node [style=none] (31) at (-4.25, 1) {};
		\node [style=none] (32) at (-4.25, 2) {};
		\node [style=none] (33) at (-4.25, 0.25) {.};
		\node [style=none] (34) at (-4.25, 0) {.};
		\node [style=none] (35) at (-4.25, -0.25) {.};
		\node [style=none] (36) at (-6.25, -0.25) {.};
		\node [style=none] (37) at (-6.25, 0) {.};
		\node [style=none] (38) at (-6.25, 0.25) {.};
		\node [style=none] (39) at (-3, 0.5) {$\propto$};
		\node [style=none] (40) at (-4, 2) {};
		\node [style=none] (41) at (-4, 1) {};
		\node [style=none] (42) at (-4, -1) {};
		\node [style=none] (43) at (-6.5, -1) {};
		\node [style=none] (44) at (-6.5, 1) {};
		\node [style=none] (45) at (-6.5, 2) {};
	\end{pgfonlayer}
	\begin{pgfonlayer}{edgelayer}
		\draw [in=135, out=0, looseness=0.50] (3.center) to (2);
		\draw [in=-180, out=45, looseness=0.50] (2) to (7.center);
		\draw [in=-180, out=15, looseness=0.50] (2) to (6.center);
		\draw [in=165, out=0, looseness=0.75] (1.center) to (2);
		\draw [in=15, out=-165, looseness=0.50] (2) to (4.center);
		\draw [in=-180, out=-15, looseness=0.75] (2) to (5.center);
		\draw (19) to (3.center);
		\draw (18) to (1.center);
		\draw (17) to (4.center);
		\draw (16) to (5.center);
		\draw (15) to (6.center);
		\draw (14) to (7.center);
		\draw (23.center) to (19);
		\draw (24.center) to (18);
		\draw (14) to (20.center);
		\draw (15) to (21.center);
		\draw (16) to (22.center);
		\draw (17) to (25.center);
		\draw [in=135, out=0, looseness=0.50] (28.center) to (27);
		\draw [in=-180, out=45, looseness=0.50] (27) to (32.center);
		\draw [in=-180, out=15, looseness=0.50] (27) to (31.center);
		\draw [in=165, out=0, looseness=0.75] (26.center) to (27);
		\draw [in=15, out=-165, looseness=0.50] (27) to (29.center);
		\draw [in=-180, out=-15, looseness=0.75] (27) to (30.center);
		\draw (45.center) to (28.center);
		\draw (44.center) to (26.center);
		\draw (43.center) to (29.center);
		\draw (42.center) to (30.center);
		\draw (41.center) to (31.center);
		\draw (40.center) to (32.center);
	\end{pgfonlayer}
\end{tikzpicture}

%% file: figures/pi-copy.tikz
\begin{tikzpicture}
	\begin{pgfonlayer}{nodelayer}
		\node [style=none] (1) at (-0.25, 1) {};
		\node [style=white dot, minimum size=.66cm] (2) at (0.75, 0) {$-\alpha$};
		\node [style=none] (3) at (-0.25, 2) {};
		\node [style=none] (4) at (-0.25, -1) {};
		\node [style=none] (5) at (1.75, -1) {};
		\node [style=none] (6) at (1.75, 1) {};
		\node [style=none] (7) at (1.75, 2) {};
		\node [style=none] (8) at (1.75, 0.25) {.};
		\node [style=none] (9) at (1.75, 0) {.};
		\node [style=none] (10) at (1.75, -0.25) {.};
		\node [style=none] (11) at (-0.25, -0.25) {.};
		\node [style=none] (12) at (-0.25, 0) {.};
		\node [style=none] (13) at (-0.25, 0.25) {.};
		\node [style=gray dot] (14) at (2.25, 2) {$\pi$};
		\node [style=gray dot] (15) at (2.25, 1) {$\pi$};
		\node [style=gray dot] (16) at (2.25, -1) {$\pi$};
		\node [style=gray dot] (17) at (-0.75, -1) {$\pi$};
		\node [style=gray dot] (18) at (-0.75, 1) {$\pi$};
		\node [style=gray dot] (19) at (-0.75, 2) {$\pi$};
		\node [style=none] (20) at (2.75, 2) {};
		\node [style=none] (21) at (2.75, 1) {};
		\node [style=none] (22) at (2.75, -1) {};
		\node [style=none] (23) at (-2, 2) {};
		\node [style=none] (24) at (-2, 1) {};
		\node [style=none] (25) at (-2, -1) {};
		\node [style=none] (26) at (-6.25, 1) {};
		\node [style=white dot, minimum size=.66cm] (27) at (-5.25, 0) {$\alpha$};
		\node [style=none] (28) at (-6.25, 2) {};
		\node [style=none] (29) at (-6.25, -1) {};
		\node [style=none] (30) at (-4.25, -1) {};
		\node [style=none] (31) at (-4.25, 1) {};
		\node [style=none] (32) at (-4.25, 2) {};
		\node [style=none] (33) at (-4.25, 0.25) {.};
		\node [style=none] (34) at (-4.25, 0) {.};
		\node [style=none] (35) at (-4.25, -0.25) {.};
		\node [style=none] (36) at (-6.25, -0.25) {.};
		\node [style=none] (37) at (-6.25, 0) {.};
		\node [style=none] (38) at (-6.25, 0.25) {.};
		\node [style=none] (39) at (-3, 0.5) {$\propto$};
		\node [style=none] (40) at (-4, 2) {};
		\node [style=none] (41) at (-4, 1) {};
		\node [style=none] (42) at (-4, -1) {};
		\node [style=none] (43) at (-7.25, -1) {};
		\node [style=none] (44) at (-7.25, 1) {};
		\node [style=gray dot] (46) at (-1.5, 2) {};
		\node [style=gray dot] (47) at (-1.5, 2) {$\pi$};
		\node [style=none] (48) at (6.25, 1) {};
		\node [style=white dot, minimum size=.66cm] (49) at (7.25, 0) {$-\alpha$};
		\node [style=none] (50) at (6.25, 2) {};
		\node [style=none] (51) at (6.25, -1) {};
		\node [style=none] (52) at (8.25, -1) {};
		\node [style=none] (53) at (8.25, 1) {};
		\node [style=none] (54) at (8.25, 2) {};
		\node [style=none] (55) at (8.25, 0.25) {.};
		\node [style=none] (56) at (8.25, 0) {.};
		\node [style=none] (57) at (8.25, -0.25) {.};
		\node [style=none] (58) at (6.25, -0.25) {.};
		\node [style=none] (59) at (6.25, 0) {.};
		\node [style=none] (60) at (6.25, 0.25) {.};
		\node [style=gray dot] (61) at (8.75, 2) {$\pi$};
		\node [style=gray dot] (62) at (8.75, 1) {$\pi$};
		\node [style=gray dot] (63) at (8.75, -1) {$\pi$};
		\node [style=gray dot] (64) at (5.5, -1) {$\pi$};
		\node [style=gray dot] (65) at (5.5, 1) {$\pi$};
		\node [style=gray dot] (66) at (5.5, 2) {$2\pi$};
		\node [style=none] (67) at (9.25, 2) {};
		\node [style=none] (68) at (9.25, 1) {};
		\node [style=none] (69) at (9.25, -1) {};
		\node [style=none] (70) at (4.75, 2) {};
		\node [style=none] (71) at (4.75, 1) {};
		\node [style=none] (72) at (4.75, -1) {};
		\node [style=none] (73) at (3.75, 0.5) {$\propto$};
		\node [style=none] (74) at (12.75, 1) {};
		\node [style=white dot, minimum size=.66cm] (75) at (13.75, 0) {$-\alpha$};
		\node [style=none] (76) at (12.75, 2) {};
		\node [style=none] (77) at (12.75, -1) {};
		\node [style=none] (78) at (14.75, -1) {};
		\node [style=none] (79) at (14.75, 1) {};
		\node [style=none] (80) at (14.75, 2) {};
		\node [style=none] (81) at (14.75, 0.25) {.};
		\node [style=none] (82) at (14.75, 0) {.};
		\node [style=none] (83) at (14.75, -0.25) {.};
		\node [style=none] (84) at (12.75, -0.25) {.};
		\node [style=none] (85) at (12.75, 0) {.};
		\node [style=none] (86) at (12.75, 0.25) {.};
		\node [style=gray dot] (87) at (15.25, 2) {$\pi$};
		\node [style=gray dot] (88) at (15.25, 1) {$\pi$};
		\node [style=gray dot] (89) at (15.25, -1) {$\pi$};
		\node [style=gray dot] (90) at (12, -1) {$\pi$};
		\node [style=gray dot] (91) at (12, 1) {$\pi$};
		\node [style=gray dot] (92) at (12, 2) {};
		\node [style=none] (93) at (15.75, 2) {};
		\node [style=none] (94) at (15.75, 1) {};
		\node [style=none] (95) at (15.75, -1) {};
		\node [style=none] (96) at (11.25, 2) {};
		\node [style=none] (97) at (11.25, 1) {};
		\node [style=none] (98) at (11.25, -1) {};
		\node [style=none] (99) at (10.25, 0.5) {$\propto$};
		\node [style=none] (100) at (19.25, 1) {};
		\node [style=white dot, minimum size=.66cm] (101) at (20.25, 0) {$-\alpha$};
		\node [style=none] (102) at (19.25, 2) {};
		\node [style=none] (103) at (19.25, -1) {};
		\node [style=none] (104) at (21.25, -1) {};
		\node [style=none] (105) at (21.25, 1) {};
		\node [style=none] (106) at (21.25, 2) {};
		\node [style=none] (107) at (21.25, 0.25) {.};
		\node [style=none] (108) at (21.25, 0) {.};
		\node [style=none] (109) at (21.25, -0.25) {.};
		\node [style=none] (110) at (19.25, -0.25) {.};
		\node [style=none] (111) at (19.25, 0) {.};
		\node [style=none] (112) at (19.25, 0.25) {.};
		\node [style=gray dot] (113) at (21.75, 2) {$\pi$};
		\node [style=gray dot] (114) at (21.75, 1) {$\pi$};
		\node [style=gray dot] (115) at (21.75, -1) {$\pi$};
		\node [style=gray dot] (116) at (18.5, -1) {$\pi$};
		\node [style=gray dot] (117) at (18.5, 1) {$\pi$};
		\node [style=none] (119) at (22.25, 2) {};
		\node [style=none] (120) at (22.25, 1) {};
		\node [style=none] (121) at (22.25, -1) {};
		\node [style=none] (122) at (17.75, 2) {};
		\node [style=none] (123) at (17.75, 1) {};
		\node [style=none] (124) at (17.75, -1) {};
		\node [style=none] (125) at (16.75, 0.5) {$\propto$};
		\node [style=gray dot] (126) at (-6.75, 2) {$\pi$};
		\node [style=none] (127) at (-7.25, 2) {};
	\end{pgfonlayer}
	\begin{pgfonlayer}{edgelayer}
		\draw [in=135, out=0, looseness=0.50] (3.center) to (2);
		\draw [in=-180, out=45, looseness=0.50] (2) to (7.center);
		\draw [in=-180, out=15, looseness=0.50] (2) to (6.center);
		\draw [in=165, out=0, looseness=0.75] (1.center) to (2);
		\draw [in=15, out=-165, looseness=0.50] (2) to (4.center);
		\draw [in=-180, out=-15, looseness=0.75] (2) to (5.center);
		\draw (19) to (3.center);
		\draw (18) to (1.center);
		\draw (17) to (4.center);
		\draw (16) to (5.center);
		\draw (15) to (6.center);
		\draw (14) to (7.center);
		\draw (24.center) to (18);
		\draw (14) to (20.center);
		\draw (15) to (21.center);
		\draw (16) to (22.center);
		\draw (17) to (25.center);
		\draw [in=135, out=0, looseness=0.50] (28.center) to (27);
		\draw [in=-180, out=45, looseness=0.50] (27) to (32.center);
		\draw [in=-180, out=15, looseness=0.50] (27) to (31.center);
		\draw [in=165, out=0, looseness=0.75] (26.center) to (27);
		\draw [in=15, out=-165, looseness=0.50] (27) to (29.center);
		\draw [in=-180, out=-15, looseness=0.75] (27) to (30.center);
		\draw (44.center) to (26.center);
		\draw (43.center) to (29.center);
		\draw (42.center) to (30.center);
		\draw (41.center) to (31.center);
		\draw (40.center) to (32.center);
		\draw (23.center) to (47);
		\draw (47) to (19);
		\draw [in=135, out=0, looseness=0.50] (50.center) to (49);
		\draw [in=-180, out=45, looseness=0.50] (49) to (54.center);
		\draw [in=-180, out=15, looseness=0.50] (49) to (53.center);
		\draw [in=165, out=0, looseness=0.75] (48.center) to (49);
		\draw [in=15, out=-165, looseness=0.50] (49) to (51.center);
		\draw [in=-180, out=-15, looseness=0.75] (49) to (52.center);
		\draw (66) to (50.center);
		\draw (65) to (48.center);
		\draw (64) to (51.center);
		\draw (63) to (52.center);
		\draw (62) to (53.center);
		\draw (61) to (54.center);
		\draw (71.center) to (65);
		\draw (61) to (67.center);
		\draw (62) to (68.center);
		\draw (63) to (69.center);
		\draw (64) to (72.center);
		\draw (70.center) to (66);
		\draw [in=135, out=0, looseness=0.50] (76.center) to (75);
		\draw [in=-180, out=45, looseness=0.50] (75) to (80.center);
		\draw [in=-180, out=15, looseness=0.50] (75) to (79.center);
		\draw [in=165, out=0, looseness=0.75] (74.center) to (75);
		\draw [in=15, out=-165, looseness=0.50] (75) to (77.center);
		\draw [in=-180, out=-15, looseness=0.75] (75) to (78.center);
		\draw (92) to (76.center);
		\draw (91) to (74.center);
		\draw (90) to (77.center);
		\draw (89) to (78.center);
		\draw (88) to (79.center);
		\draw (87) to (80.center);
		\draw (97.center) to (91);
		\draw (87) to (93.center);
		\draw (88) to (94.center);
		\draw (89) to (95.center);
		\draw (90) to (98.center);
		\draw (96.center) to (92);
		\draw [in=135, out=0, looseness=0.50] (102.center) to (101);
		\draw [in=-180, out=45, looseness=0.50] (101) to (106.center);
		\draw [in=-180, out=15, looseness=0.50] (101) to (105.center);
		\draw [in=165, out=0, looseness=0.75] (100.center) to (101);
		\draw [in=15, out=-165, looseness=0.50] (101) to (103.center);
		\draw [in=-180, out=-15, looseness=0.75] (101) to (104.center);
		\draw (117) to (100.center);
		\draw (116) to (103.center);
		\draw (115) to (104.center);
		\draw (114) to (105.center);
		\draw (113) to (106.center);
		\draw (123.center) to (117);
		\draw (113) to (119.center);
		\draw (114) to (120.center);
		\draw (115) to (121.center);
		\draw (116) to (124.center);
		\draw (102.center) to (122.center);
		\draw (127.center) to (126);
		\draw (126) to (28.center);
	\end{pgfonlayer}
\end{tikzpicture}

%% file: figures/spider-fusion.tikz
\begin{tikzpicture}
	\begin{pgfonlayer}{nodelayer}
		\node [style=white dot] (0) at (-1, 1) {$\alpha$};
		\node [style=white dot] (1) at (1, -1) {$\beta$};
		\node [style=none] (3) at (2, -1.25) {};
		\node [style=none] (5) at (2, -2) {};
		\node [style=none] (7) at (-2, 2) {};
		\node [style=none] (8) at (-2, 1.25) {};
		\node [style=none] (10) at (0, 1.25) {};
		\node [style=none] (12) at (0, 2) {};
		\node [style=none] (13) at (0, -2) {};
		\node [style=none] (15) at (0, -1.25) {};
		\node [style=none] (17) at (0, 1.5) {.};
		\node [style=none] (21) at (0, 0) {..};
		\node [style=none] (22) at (0, 1.75) {};
		\node [style=none] (23) at (0, 1.75) {.};
		\node [style=none] (24) at (-2, 1.5) {.};
		\node [style=none] (26) at (-2, 1.75) {.};
		\node [style=none] (27) at (0, -1.75) {};
		\node [style=none] (28) at (0, -1.75) {.};
		\node [style=none] (29) at (0, -1.5) {.};
		\node [style=none] (31) at (2, -1.75) {.};
		\node [style=none] (34) at (2, -1.5) {.};
		\node [style=none] (35) at (3, 0) {$\propto$};
		\node [style=white dot] (37) at (5.5, 0) {\tiny{}$\alpha + \beta$};
		\node [style=none] (39) at (4, 1) {};
		\node [style=none] (40) at (4, 0.25) {};
		\node [style=none] (41) at (4, 0.5) {.};
		\node [style=none] (42) at (4, 0.75) {.};
		\node [style=none] (43) at (4, -0.25) {};
		\node [style=none] (44) at (4, -1) {};
		\node [style=none] (45) at (4, -0.75) {.};
		\node [style=none] (46) at (4, -0.5) {.};
		\node [style=none] (47) at (7, 1) {};
		\node [style=none] (48) at (7, 0.25) {};
		\node [style=none] (49) at (7, 0.5) {.};
		\node [style=none] (50) at (7, 0.75) {.};
		\node [style=none] (51) at (7, -0.25) {};
		\node [style=none] (52) at (7, -1) {};
		\node [style=none] (53) at (7, -0.75) {.};
		\node [style=none] (54) at (7, -0.5) {.};
	\end{pgfonlayer}
	\begin{pgfonlayer}{edgelayer}
		\draw [in=120, out=-30] (0) to (1);
		\draw [in=150, out=-60] (0) to (1);
		\draw [in=-180, out=0, looseness=1.25] (1) to (3.center);
		\draw [in=180, out=-30] (1) to (5.center);
		\draw [in=0, out=-180] (1) to (15.center);
		\draw [in=0, out=-135] (1) to (13.center);
		\draw [in=180, out=0, looseness=1.25] (0) to (10.center);
		\draw [in=180, out=30] (0) to (12.center);
		\draw [in=0, out=135, looseness=1.25] (0) to (7.center);
		\draw [in=0, out=165] (0) to (8.center);
		\draw [in=-180, out=30] (37) to (47.center);
		\draw [in=-180, out=15] (37) to (48.center);
		\draw [in=180, out=-15] (37) to (51.center);
		\draw [in=180, out=-30] (37) to (52.center);
		\draw [in=0, out=-150] (37) to (44.center);
		\draw [in=0, out=-165, looseness=1.25] (37) to (43.center);
		\draw [in=0, out=150, looseness=0.75] (37) to (39.center);
		\draw [in=0, out=165] (37) to (40.center);
	\end{pgfonlayer}
\end{tikzpicture}

%% file: figures/spider-fusion-1-1.tikz
\begin{tikzpicture}
	\begin{pgfonlayer}{nodelayer}
		\node [style=white dot] (0) at (-1.5, 0) {$\alpha$};
		\node [style=white dot] (1) at (0.5, 0) {$\beta$};
		\node [style=none] (3) at (2, 1) {};
		\node [style=none] (5) at (2, -1) {};
		\node [style=none] (31) at (2, 0) {.};
		\node [style=none] (34) at (2, 0.25) {.};
		\node [style=none] (35) at (3, 0) {$\propto$};
		\node [style=white dot] (37) at (6, 0) {\tiny{}$\alpha + \beta$};
		\node [style=none] (55) at (2, -0.25) {.};
		\node [style=none] (56) at (-3, 1) {};
		\node [style=none] (57) at (-3, -1) {};
		\node [style=none] (58) at (-3, 0) {.};
		\node [style=none] (59) at (-3, 0.25) {.};
		\node [style=none] (60) at (-3, -0.25) {.};
		\node [style=none] (61) at (4, 1) {};
		\node [style=none] (62) at (4, -1) {};
		\node [style=none] (63) at (4, 0) {.};
		\node [style=none] (64) at (4, 0.25) {.};
		\node [style=none] (65) at (4, -0.25) {.};
		\node [style=none] (66) at (8, 1) {};
		\node [style=none] (67) at (8, -1) {};
		\node [style=none] (68) at (8, 0) {.};
		\node [style=none] (69) at (8, 0.25) {.};
		\node [style=none] (70) at (8, -0.25) {.};
	\end{pgfonlayer}
	\begin{pgfonlayer}{edgelayer}
		\draw [in=-180, out=15, looseness=1.25] (1) to (3.center);
		\draw [in=180, out=-15, looseness=1.25] (1) to (5.center);
		\draw [in=165, out=0, looseness=1.25] (56.center) to (0);
		\draw [in=0, out=-165, looseness=1.25] (0) to (57.center);
		\draw (0) to (1);
		\draw [in=-180, out=15, looseness=0.75] (37) to (66.center);
		\draw [in=180, out=-15, looseness=0.75] (37) to (67.center);
		\draw [in=0, out=-165, looseness=0.75] (37) to (62.center);
		\draw [in=0, out=165, looseness=0.75] (37) to (61.center);
	\end{pgfonlayer}
\end{tikzpicture}

%% file: figures/shift-qubits-small.tikz
\begin{tikzpicture}
	\begin{pgfonlayer}{nodelayer}
		\node [style=none] (0) at (-7, 2) {};
		\node [style=none] (1) at (-4, 2) {};
		\node [style=none] (2) at (-7, 1) {};
		\node [style=none] (3) at (-4, 1) {};
		\node [style=none] (4) at (-7, 0) {};
		\node [style=none] (5) at (-4, -1) {};
		\node [style=none] (6) at (-7, -2) {};
		\node [style=none] (7) at (-4, -2) {};
		\node [style=none] (8) at (-4, 0) {};
		\node [style=none] (9) at (-4, 0) {};
		\node [style=none] (10) at (-4, 0) {.};
		\node [style=none] (11) at (-4, -0.25) {};
		\node [style=none] (12) at (-4, -0.25) {.};
		\node [style=none] (13) at (-4, 0.25) {};
		\node [style=none] (14) at (-4, 0.25) {};
		\node [style=none] (15) at (-4, 0.25) {.};
		\node [style=none] (16) at (-7, -1) {};
		\node [style=none] (17) at (-7, -1) {};
		\node [style=none] (18) at (-7, -1) {.};
		\node [style=none] (19) at (-7, -1.25) {};
		\node [style=none] (20) at (-7, -1.25) {.};
		\node [style=none] (21) at (-7, -0.75) {};
		\node [style=none] (22) at (-7, -0.75) {};
		\node [style=none] (23) at (-7, -0.75) {.};
		\node [style=none] (24) at (-2, 2) {};
		\node [style=none] (25) at (0, 1) {};
		\node [style=none] (26) at (-2, 1) {};
		\node [style=none] (27) at (-2, 1) {};
		\node [style=none] (28) at (0, 2) {};
		\node [style=none] (29) at (0, 0) {};
		\node [style=none] (30) at (2, 0) {};
		\node [style=none] (31) at (2, 1) {};
		\node [style=none] (32) at (-2, 0) {};
		\node [style=none] (33) at (-2, -2) {};
		\node [style=none] (34) at (2, -2) {};
		\node [style=none] (35) at (-2, -1) {};
		\node [style=none] (36) at (-2, -1) {};
		\node [style=none] (37) at (-2, -1) {.};
		\node [style=none] (38) at (-2, -1.25) {};
		\node [style=none] (39) at (-2, -1.25) {.};
		\node [style=none] (40) at (-2, -0.75) {};
		\node [style=none] (41) at (-2, -0.75) {};
		\node [style=none] (42) at (-2, -0.75) {.};
		\node [style=none] (43) at (4, -1) {};
		\node [style=none] (44) at (6, -2) {};
		\node [style=none] (45) at (6, -1) {};
		\node [style=none] (46) at (4, -2) {};
		\node [style=none] (48) at (2.75, -0.5) {.};
		\node [style=none] (53) at (3.25, -0.75) {.};
		\node [style=none] (54) at (4, 1) {};
		\node [style=none] (55) at (6, 1) {};
		\node [style=none] (56) at (2, 2) {};
		\node [style=none] (57) at (4, 2) {};
		\node [style=none] (58) at (6, 2) {};
		\node [style=none] (59) at (2.75, -2) {.};
		\node [style=none] (60) at (3.25, -2) {.};
		\node [style=none] (61) at (2.75, 1) {.};
		\node [style=none] (62) at (3.25, 1) {.};
		\node [style=none] (63) at (2.75, 2) {.};
		\node [style=none] (64) at (3.25, 2) {.};
		\node [style=none] (66) at (-3, 0) {$\propto$};
		\node [style=none] (129) at (2, -1) {};
		\node [style=none] (130) at (2, -1) {};
		\node [style=none] (131) at (2, -1) {.};
		\node [style=none] (132) at (2, -1.25) {};
		\node [style=none] (133) at (2, -1.25) {.};
		\node [style=none] (134) at (2, -0.75) {};
		\node [style=none] (135) at (2, -0.75) {};
		\node [style=none] (136) at (2, -0.75) {.};
	\end{pgfonlayer}
	\begin{pgfonlayer}{edgelayer}
		\draw [in=-180, out=0, looseness=0.50] (0.center) to (7.center);
		\draw [in=180, out=0, looseness=0.50] (6.center) to (5.center);
		\draw [in=-180, out=0, looseness=0.50] (4.center) to (3.center);
		\draw [in=180, out=0, looseness=0.50] (2.center) to (1.center);
		\draw [in=-180, out=0, looseness=0.75] (24.center) to (25.center);
		\draw [in=-180, out=0, looseness=0.75] (27.center) to (28.center);
		\draw (32.center) to (29.center);
		\draw [in=-180, out=0, looseness=0.75] (29.center) to (31.center);
		\draw [in=-180, out=0, looseness=0.75] (25.center) to (30.center);
		\draw (33.center) to (34.center);
		\draw [in=-180, out=0, looseness=0.75] (46.center) to (45.center);
		\draw [in=180, out=0, looseness=0.75] (43.center) to (44.center);
		\draw (57.center) to (58.center);
		\draw (55.center) to (54.center);
		\draw (28.center) to (56.center);
	\end{pgfonlayer}
\end{tikzpicture}

%% file: figures/ingest-rotation.tikz
\begin{tikzpicture}
	\begin{pgfonlayer}{nodelayer}
		\node [style=white dot] (0) at (8, 0) {$x$};
		\node [style=gray dot] (1) at (6, 0) {$z$};
		\node [style=hadamard] (2) at (0, 0) {Y};
		\node [style=hadamard] (3) at (4, 0) {Y};
		\node [style=gray dot] (4) at (2, 0) {$y$};
		\node [style=none] (5) at (-2, 0) {};
		\node [style=none] (6) at (9.5, 0) {};
	\end{pgfonlayer}
	\begin{pgfonlayer}{edgelayer}
		\draw (2) to (4);
		\draw (4) to (3);
		\draw (5.center) to (2);
		\draw (3) to (1);
		\draw (1) to (0);
		\draw (0) to (6.center);
	\end{pgfonlayer}
\end{tikzpicture}

%% file: figures/ingest-cnot.tikz
\begin{tikzpicture}
	\begin{pgfonlayer}{nodelayer}
		\node [style=none] (1) at (-5, 1) {};
		\node [style=none] (2) at (-5, 0) {};
		\node [style=none] (4) at (-5, -4) {};
		\node [style=none] (5) at (-3, -4) {};
		\node [style=none] (7) at (-3, 0) {};
		\node [style=none] (8) at (-3, 1) {};
		\node [style=none] (10) at (-3, 3) {};
		\node [style=none] (11) at (-5, 3) {};
		\node [style=none] (12) at (-1, 0) {};
		\node [style=none] (13) at (-1, 1) {};
		\node [style=none] (15) at (-1, 3) {};
		\node [style=none] (16) at (-1, -4) {};
		\node [style=none] (17) at (1, -4) {};
		\node [style=none] (18) at (1, 1) {};
		\node [style=none] (19) at (1, 0) {};
		\node [style=none] (21) at (1, 3) {};
		\node [style=none] (25) at (1, 2) {};
		\node [style=none] (26) at (1, 2) {};
		\node [style=none] (27) at (1, 2) {.};
		\node [style=none] (28) at (1, 1.75) {};
		\node [style=none] (29) at (1, 1.75) {.};
		\node [style=none] (30) at (1, 2.25) {};
		\node [style=none] (31) at (1, 2.25) {};
		\node [style=none] (32) at (1, 2.25) {.};
		\node [style=none] (33) at (-5, 2) {};
		\node [style=none] (34) at (-5, 2) {};
		\node [style=none] (35) at (-5, 2) {.};
		\node [style=none] (36) at (-5, 1.75) {};
		\node [style=none] (37) at (-5, 1.75) {.};
		\node [style=none] (38) at (-5, 2.25) {};
		\node [style=none] (39) at (-5, 2.25) {};
		\node [style=none] (40) at (-5, 2.25) {.};
		\node [style=none] (41) at (-5, -1) {};
		\node [style=none] (42) at (-5, -3) {};
		\node [style=none] (43) at (-3, -3) {};
		\node [style=none] (44) at (-1, -3) {};
		\node [style=none] (45) at (1, -3) {};
		\node [style=none] (46) at (1, -1) {};
		\node [style=none] (47) at (-1, -1) {};
		\node [style=none] (48) at (-3, -1) {};
		\node [style=none] (49) at (-5, -7) {};
		\node [style=none] (50) at (-3, -7) {};
		\node [style=none] (51) at (-3, -5) {};
		\node [style=none] (52) at (-5, -5) {};
		\node [style=none] (53) at (-1, -7) {};
		\node [style=none] (54) at (-1, -5) {};
		\node [style=none] (55) at (1, -7) {};
		\node [style=none] (56) at (1, -5) {};
		\node [style=none] (57) at (1, -6) {};
		\node [style=none] (58) at (1, -6) {};
		\node [style=none] (59) at (1, -6) {.};
		\node [style=none] (60) at (1, -6.25) {};
		\node [style=none] (61) at (1, -6.25) {.};
		\node [style=none] (62) at (1, -5.75) {};
		\node [style=none] (63) at (1, -5.75) {};
		\node [style=none] (64) at (1, -5.75) {.};
		\node [style=none] (65) at (-5, -6) {};
		\node [style=none] (66) at (-5, -6) {};
		\node [style=none] (67) at (-5, -6) {.};
		\node [style=none] (68) at (-5, -6.25) {};
		\node [style=none] (69) at (-5, -6.25) {.};
		\node [style=none] (70) at (-5, -5.75) {};
		\node [style=none] (71) at (-5, -5.75) {};
		\node [style=none] (72) at (-5, -5.75) {.};
		\node [style=none] (73) at (1, -2) {};
		\node [style=none] (74) at (1, -2) {};
		\node [style=none] (75) at (1, -2) {.};
		\node [style=none] (76) at (1, -2.25) {};
		\node [style=none] (77) at (1, -2.25) {.};
		\node [style=none] (78) at (1, -1.75) {};
		\node [style=none] (79) at (1, -1.75) {};
		\node [style=none] (80) at (1, -1.75) {.};
		\node [style=none] (81) at (-5, -2) {};
		\node [style=none] (82) at (-5, -2) {};
		\node [style=none] (83) at (-5, -2) {.};
		\node [style=none] (84) at (-5, -2.25) {};
		\node [style=none] (85) at (-5, -2.25) {.};
		\node [style=none] (86) at (-5, -1.75) {};
		\node [style=none] (87) at (-5, -1.75) {};
		\node [style=none] (88) at (-5, -1.75) {.};
		\node [style=gray dot] (89) at (-2, 1) {};
		\node [style=white dot] (90) at (-2, 0) {};
		\node [style=none] (91) at (2, 3) {};
		\node [style=none] (92) at (2, 1) {};
		\node [style=none] (93) at (2, 0) {};
		\node [style=none] (94) at (2, -1) {};
		\node [style=none] (95) at (2, -3) {};
		\node [style=none] (96) at (2, -4) {};
		\node [style=none] (97) at (2, -5) {};
		\node [style=none] (98) at (2, -7) {};
		\node [style=none] (99) at (-6, -7) {};
		\node [style=none] (100) at (-6, -5) {};
		\node [style=none] (101) at (-6, -4) {};
		\node [style=none] (102) at (-6, -3) {};
		\node [style=none] (103) at (-6, -1) {};
		\node [style=none] (104) at (-6, 0) {};
		\node [style=none] (105) at (-6, 1) {};
		\node [style=none] (106) at (-6, 3) {};
		\node [style=none] (107) at (-7, 1) {$n$};
		\node [style=none] (108) at (-7, 0) {$n+1$};
		\node [style=none] (109) at (-7, 3) {$1$};
		\node [style=none] (110) at (-7, -4) {$c$};
		\node [style=none] (111) at (-7, -3) {$c-1$};
		\node [style=none] (112) at (-7, -1) {$n+2$};
		\node [style=none] (113) at (-7, -5) {$c+1$};
		\node [style=none] (114) at (-7, -7) {$q$};
	\end{pgfonlayer}
	\begin{pgfonlayer}{edgelayer}
		\draw [in=-180, out=0, looseness=0.75] (4.center) to (7.center);
		\draw [in=180, out=0, looseness=0.75] (2.center) to (5.center);
		\draw (1.center) to (8.center);
		\draw (10.center) to (11.center);
		\draw (15.center) to (10.center);
		\draw (13.center) to (18.center);
		\draw (21.center) to (15.center);
		\draw (41.center) to (48.center);
		\draw (48.center) to (47.center);
		\draw (47.center) to (46.center);
		\draw (45.center) to (42.center);
		\draw [in=180, out=0, looseness=0.75] (16.center) to (19.center);
		\draw [in=180, out=0, looseness=0.75] (12.center) to (17.center);
		\draw (49.center) to (50.center);
		\draw (51.center) to (52.center);
		\draw (54.center) to (51.center);
		\draw (50.center) to (53.center);
		\draw (53.center) to (55.center);
		\draw (56.center) to (54.center);
		\draw (8.center) to (89);
		\draw (89) to (13.center);
		\draw (12.center) to (90);
		\draw (90) to (7.center);
		\draw (90) to (89);
		\draw (5.center) to (16.center);
		\draw (11.center) to (106.center);
		\draw (21.center) to (91.center);
		\draw (18.center) to (92.center);
		\draw (19.center) to (93.center);
		\draw (46.center) to (94.center);
		\draw (45.center) to (95.center);
		\draw (17.center) to (96.center);
		\draw (97.center) to (56.center);
		\draw (98.center) to (55.center);
		\draw (49.center) to (99.center);
		\draw (52.center) to (100.center);
		\draw (42.center) to (102.center);
		\draw (101.center) to (4.center);
		\draw (41.center) to (103.center);
		\draw (104.center) to (2.center);
		\draw (105.center) to (1.center);
	\end{pgfonlayer}
\end{tikzpicture}

%% file: figures/deformation.tikz
\begin{tikzpicture}
	\begin{pgfonlayer}{nodelayer}
		\node [style=white dot] (0) at (-2, 3) {};
		\node [style=gray dot] (1) at (-2, 2) {};
		\node [style=gray dot] (2) at (-2, 4) {};
		\node [style=white dot] (3) at (1, 2) {};
		\node [style=white dot] (4) at (1, 4) {};
		\node [style=none] (5) at (-3, 4) {};
		\node [style=none] (6) at (-3, 3) {};
		\node [style=none] (7) at (-3, 2) {};
		\node [style=gray dot] (8) at (4, 4) {};
		\node [style=gray dot] (10) at (4, 2) {};
		\node [style=none] (11) at (5, 4) {};
		\node [style=none] (13) at (5, 2) {};
		\node [style=white dot] (14) at (8, 2) {};
		\node [style=gray dot] (15) at (8, 3) {};
		\node [style=gray dot] (16) at (8, 4) {};
		\node [style=white dot] (17) at (13, 3) {};
		\node [style=white dot] (18) at (11, 4) {};
		\node [style=none] (19) at (7, 4) {};
		\node [style=none] (20) at (7, 3) {};
		\node [style=none] (21) at (7, 2) {};
		\node [style=gray dot] (22) at (14, 4) {};
		\node [style=gray dot] (23) at (14, 3) {};
		\node [style=none] (24) at (15, 4) {};
		\node [style=none] (25) at (15, 2) {};
		\node [style=none] (26) at (6, 3) {$=$};
	\end{pgfonlayer}
	\begin{pgfonlayer}{edgelayer}
		\draw (2) to (4);
		\draw [in=195, out=15] (0) to (4);
		\draw [in=165, out=-15] (0) to (3);
		\draw (1) to (3);
		\draw (4) to (8);
		\draw (10) to (3);
		\draw [bend left=75, looseness=1.25] (4) to (3);
		\draw (8) to (11.center);
		\draw (10) to (13.center);
		\draw (6.center) to (0);
		\draw (5.center) to (2);
		\draw (1) to (7.center);
		\draw (16) to (18);
		\draw [in=-165, out=15] (14) to (18);
		\draw [in=-165, out=0] (14) to (17);
		\draw (15) to (17);
		\draw (18) to (22);
		\draw (23) to (17);
		\draw [in=15, out=-15] (18) to (17);
		\draw [in=180, out=0] (22) to (24.center);
		\draw [in=165, out=-15, looseness=1.50] (23) to (25.center);
		\draw [in=-180, out=0, looseness=0.75] (20.center) to (14);
		\draw (19.center) to (16);
		\draw [in=0, out=-180, looseness=0.75] (15) to (21.center);
	\end{pgfonlayer}
\end{tikzpicture}

%% file: sections/future.tex
\section{Graph Representation}\label{sec:graph}

\begin{wrapfigure}{R}{0.5\textwidth}
  \centering
  \input{annotation-alg.tikz}
  \caption{\centering An example of the edge annotation algorithm. Each spider is shown with a unique node number inside of it. In the first step, all wires out of the spiders are annotated with the node numbers of the spiders. Then when stacking nodes, those numbers are carried forward and applied to the stack constructions wires. Note that the order is preserved. We see that the outer set of numbers is carried forward upon a composition in the last step. As before, \(\updownarrow\) denotes stacking and \(\leftrightarrow\) composition}\label{fig:annotation-alg}
\end{wrapfigure}
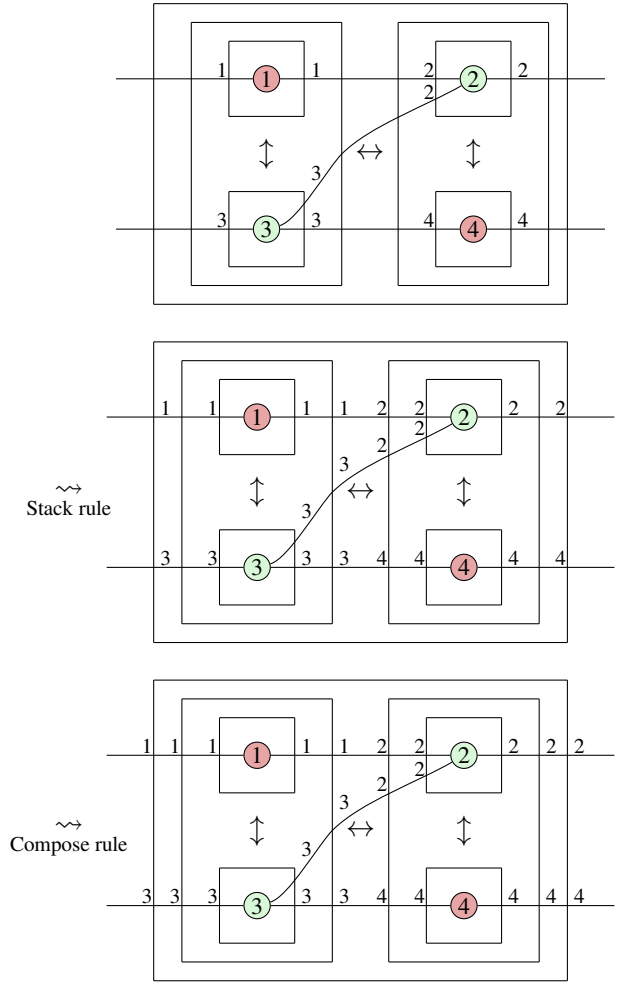

As discussed in \Cref{sec:blk}, we can view diagrams in both block representation and graph representation.
In graph representation, we construct ZX diagrams solely based on node adjacency.
This representation brings some interesting properties that will allow us to optimize diagrams or prove further rules (as mentioned in \Cref{sec:rules}).
We can see that any graph representation diagram can be deformed arbitrarily, as long as inputs and outputs are kept in order.
\Cref{fig:deform} illustrates this: Here, we see two ZX diagrams equal up to deformation.

One of our future goals is to add a graph representation for ZX diagrams, allowing us to act on diagrams.%
Creating a verified semantics for graph representation remains a work-in-progress since, fundamentally, this requires topologically-sorted graph traversal, which is a hard problem to implement in proof assistants.
Further, converting graph representation diagrams into block representation is also challenging; if overcome, we could provide semantics to graph representation in terms of the corresponding block representation diagrams. 
We devised a method to convert to graph representation, which is a central problem.
In the following, we shall outline this conversion.

Before converting into graph representation, we convert ZX diagrams into a restricted yet just as universal form; this form is similar to Duncan et al's~\cite{duncan-et-al-2020} graph-like form (as described in \Cref{sec:zx-opt}), differing only in the existence of self-loops and two or more possible links between spiders.
The restrictions are as follows:

\begin{enumerate}[nolistsep]
  \item Only Hadamard edges
  \item Only one type of spider (Z spiders)
  \item Every spider has one input and zero to two outputs or vice-versa
\end{enumerate}

Restrictions 1 \& 2 are in place to make our graph more conventional (and akin to graph-like diagrams) by having nodes and edges.
Restriction 3, however, is in place for ease of proving.
To divide up the proofs into logical steps, there are separate intermediate representations that build up all three restrictions.
These IRs exist to divide proofs into logical components and are mostly transparent.
Developers, however, can choose to use less restricted forms.

Given our restricted form, our algorithm proceeds as follows:
A procedure numbers all ZX diagram components (i.e. the constructors)
with a unique integer.
Then it proceeds to number edges of components as follows:
Each component with \(n\) inputs and \(m\) outputs will produce a pair of lists sized \(n\) and \(m\), where every position in the list indicates the closest fundamental component (spider or cap/cup).
A non stacking/composing component with \(n\) inputs, \(m\) outputs, and node number \(x\) will return 
lists of size $n$ and $m$, each with all entries being $x$.

When stacking two diagrams, the procedure concatenates respective lists, and when sequencing diagrams, it carries forward the outer lists.
\Cref{fig:annotation-alg} shows an example of this process.
It is important to note that we treat caps and cups like spiders at this stage.
Once the edge numbering is complete, we will traverse the diagram once more, and at every Compose, we will match the output edge numbers of the left diagram with the input edge numbers of the right diagram and mark each of those as an edge.
Looking at our example in \Cref{fig:annotation-alg}, we see wires labeled \((1,2)\), \((3,2)\), and \((3,4)\) bridging the main composition. 
We use the information from the algorithm to infer which inputs/outputs of the diagram are connected to which node by looking at the outermost annotation.
So in our example, we see by looking at the outermost labels that input 1 is connected to node 1, input 2 to node 3, output 1 to node 2, and output 2 to node 4.
We can then annotate the entire diagram's inputs and outputs by looking at the outermost labels.
\Cref{alg:blk-graph} shows a pseudocode description of these processes.

\begin{algorithm}[tb]
\scriptsize
 \caption{block representation to graph representation conversion}\label{alg:blk-graph}
  \begin{algorithmic}
    \Function{NumberNodes}{zx}
      \State assignFreshNumber(zx)
      \If{zx = Stack zx1 zx2 OR zx = Compose zx1 zx2}
        \State NumberInnerNodes(zx1)
        \State NumberInnerNodes(zx2)
      \EndIf
    \EndFunction

  \Function{NumberEdges}{zx}
    \If{zx = Stack zx1 zx2}
      \State (in1, out1) = NumberEdges(zx1)
      \State (in2, out2) = NumberEdges(zx2)
      \State \Return (in1 ++ in2), (out1 ++ out2)
    \ElsIf{zx = Compose zx1 zx2}
      \State (in1, \_) = NumberEdges(zx1)
      \State (\_, out2) = NumberEdges(zx2)
      \State \Return in1, out2
    \Else
      \State \Return NodeNumber(zx), NodeNumber(zx)
    \EndIf
  \EndFunction

  \Function{CreateEdges}{zx}
    \If{zx = Compose zx1 zx2}
      \State (\_, out1) = GetEdgeNumbers(zx1)
      \State (in2, \_) = GetEdgeNumbers(zx2)
      
      \For{((in, out) \( \in \) (out1, in2))}
        \Comment Note that out1 and in2 have the same length by construction
        \State AddEdge(in, out) 
      \EndFor
    \EndIf
    \If{zx = Compose zx1 zx2 OR Stack zx1 zx2}
      \State CreateEdges(zx1)
      \State CreateEdges(zx2)
    \EndIf
  \EndFunction
  \end{algorithmic}
\end{algorithm}

\section{Future Directions}\label{sec:future}

\subsection{Circuit extraction}

Once we have converted our block representation diagrams to graph representation diagrams, we will extract these graph representation diagrams to \SQIR circuits.
To accomplish this, we plan to follow Backens extraction work \cite{backens2021there}.
Our systems should allow us to define a notion of \textit{gflow}. 
This graph-theoretical property is sufficient for extracting ZX diagrams into circuits,  which will be a valuable tool to verify that optimizations do not break extractability.
Circuit extraction will complete the core of \VyZX as now we will be able to ingest circuits, write functions over them, and extract the circuits back to \SQIR.

Once we have verified optimizations and extractions, we are able to pursue a couple of interesting projects, including optimization and simulation.
As we are sharing base libraries with \VOQC, the natural next idea is to integrate \VyZX with \VOQC properly.

\subsection{\VOQC integration}

Once we complete graph representation conversion and circuit extraction, we plan to build an optimization pass using \pyZX-like optimizations, as described in \Cref{sec:zx-opt}.
Instead of building a standalone optimizer, we plan on integrating our work into \VOQC. 
Since we can ingest from (and later extract to) \SQIR, we have a common IR that will allow us to have a wholly verified pipeline and interoperate with ease.
It will be interesting to see whether non-ZX optimizations combined with ZX optimizations yield a benefit.
\VOQC has the advantage of a complete interface that allows for pass selection, allowing us to expose the ZX-based optimizations to users.
Furthermore, integration into \VOQC will allow us to benchmark our optimizer against other state-of-the-art optimizers like \pyZX~\cite{kissinger2020Pyzx} and \quartz~\cite{xu2022}.%

\subsection{The ZH and the ZW Calculi}
\VyZX's design is focused exclusively on the ZX calculus, but the principles described here could easily be applied to other similar calculi such as the ZW calculus~\cite{hadzihasanovic2015diagrammatic} or ZH calculus~\cite{backens2019zh}, which have broad applications to quantum communication the description of quantum oracles, respectively.
In fact, block construct diagrams let us easily translate between various graphical calculi.
The translation can be verified by Coq by a simple inductive proof over the signature for the calculus and our shared string diagram constructions.
It may be of interest to develop additional optimizations based on these different calculi, and a small extension to \VyZX could allow such optimizations to be verified.
With the different IRs we have right now, we are confident that extensions of the calculus are easy to integrate.

%% file: figures/annotation-alg.tikz
\begin{tikzpicture}
	\begin{pgfonlayer}{nodelayer}
		\node [style=none] (0) at (17, 3.5) {};
		\node [style=white dot] (1) at (18, 2.5) {1};
		\node [style=none] (2) at (19, 3.5) {};
		\node [style=none] (3) at (19, 1.5) {};
		\node [style=none] (4) at (17, 1.5) {};
		\node [style=none] (5) at (17, -0.5) {};
		\node [style=none] (6) at (19, -0.5) {};
		\node [style=none] (7) at (17, -2.5) {};
		\node [style=none] (8) at (19, -2.5) {};
		\node [style=gray dot] (9) at (18, -1.5) {3};
		\node [style=none] (10) at (20, 4) {};
		\node [style=none] (11) at (20, -3) {};
		\node [style=none] (12) at (16, -3) {};
		\node [style=none] (13) at (16, 4) {};
		\node [style=none] (15) at (18, 0.5) {$\updownarrow$};
		\node [style=none] (17) at (22.5, 3.5) {};
		\node [style=gray dot] (18) at (23.5, 2.5) {2};
		\node [style=none] (19) at (24.5, 3.5) {};
		\node [style=none] (20) at (24.5, 1.5) {};
		\node [style=none] (21) at (22.5, 1.5) {};
		\node [style=none] (22) at (22.5, -0.5) {};
		\node [style=none] (23) at (24.5, -0.5) {};
		\node [style=none] (24) at (22.5, -2.5) {};
		\node [style=none] (25) at (24.5, -2.5) {};
		\node [style=white dot] (26) at (23.5, -1.5) {4};
		\node [style=none] (27) at (25.5, 4) {};
		\node [style=none] (28) at (25.5, -3) {};
		\node [style=none] (29) at (21.5, -3) {};
		\node [style=none] (30) at (21.5, 4) {};
		\node [style=none] (31) at (23.5, 0.5) {$\updownarrow$};
		\node [style=none] (32) at (25.25, -0.5) {};
		\node [style=none] (33) at (20.75, 0.5) {$\leftrightarrow$};
		\node [style=none] (36) at (20, 0.5) {};
		\node [style=none] (37) at (27.5, 2.5) {};
		\node [style=none] (38) at (27.5, -1.5) {};
		\node [style=none] (39) at (14, -1.5) {};
		\node [style=none] (40) at (14, 2.5) {};
		\node [style=none] (41) at (19.25, 2.75) {\scriptsize{} 1};
		\node [style=none] (43) at (20.25, 2.75) {\scriptsize{} 1};
		\node [style=none] (44) at (20.25, 1.25) {\scriptsize{} 3};
		\node [style=none] (46) at (20.25, -1.25) {\scriptsize{} 3};
		\node [style=none] (47) at (19.25, -1.25) {\scriptsize{} 3};
		\node [style=none] (49) at (19.25, 0) {\scriptsize{} 3};
		\node [style=none] (50) at (16.75, -1.25) {\scriptsize{} 3};
		\node [style=none] (51) at (15.5, -1.25) {\scriptsize{} 3};
		\node [style=none] (52) at (15.5, 2.75) {\scriptsize{} 1};
		\node [style=none] (53) at (16.75, 2.75) {};
		\node [style=none] (54) at (16.75, 2.75) {\scriptsize{} 1};
		\node [style=none] (55) at (21.25, 2.75) {\scriptsize{} 2};
		\node [style=none] (56) at (21.25, 1.75) {\scriptsize{} 2};
		\node [style=none] (58) at (21.25, -1.25) {\scriptsize{} 4};
		\node [style=none] (59) at (22.25, 2.75) {\scriptsize{} 2};
		\node [style=none] (60) at (22.25, 2.2) {\scriptsize{} 2};
		\node [style=none] (62) at (22.25, -1.25) {\scriptsize{} 4};
		\node [style=none] (65) at (26, 2.75) {\scriptsize{} 2};
		\node [style=none] (66) at (26, -1.25) {};
		\node [style=none] (67) at (26, -1.25) {\scriptsize{} 4};
		\node [style=none] (68) at (17.25, 12.5) {};
		\node [style=white dot] (69) at (18.25, 11.5) {1};
		\node [style=none] (70) at (19.25, 12.5) {};
		\node [style=none] (71) at (19.25, 10.5) {};
		\node [style=none] (72) at (17.25, 10.5) {};
		\node [style=none] (73) at (17.25, 8.5) {};
		\node [style=none] (74) at (19.25, 8.5) {};
		\node [style=none] (75) at (17.25, 6.5) {};
		\node [style=none] (76) at (19.25, 6.5) {};
		\node [style=gray dot] (77) at (18.25, 7.5) {3};
		\node [style=none] (78) at (20.25, 13) {};
		\node [style=none] (79) at (20.25, 6) {};
		\node [style=none] (80) at (16.25, 6) {};
		\node [style=none] (81) at (16.25, 13) {};
		\node [style=none] (82) at (18.25, 9.5) {$\updownarrow$};
		\node [style=none] (83) at (22.75, 12.5) {};
		\node [style=gray dot] (84) at (23.75, 11.5) {2};
		\node [style=none] (85) at (24.75, 12.5) {};
		\node [style=none] (86) at (24.75, 10.5) {};
		\node [style=none] (87) at (22.75, 10.5) {};
		\node [style=none] (88) at (22.75, 8.5) {};
		\node [style=none] (89) at (24.75, 8.5) {};
		\node [style=none] (90) at (22.75, 6.5) {};
		\node [style=none] (91) at (24.75, 6.5) {};
		\node [style=white dot] (92) at (23.75, 7.5) {4};
		\node [style=none] (93) at (25.75, 13) {};
		\node [style=none] (94) at (25.75, 6) {};
		\node [style=none] (95) at (21.75, 6) {};
		\node [style=none] (96) at (21.75, 13) {};
		\node [style=none] (97) at (23.75, 9.5) {$\updownarrow$};
		\node [style=none] (98) at (25.5, 8.5) {};
		\node [style=none] (99) at (21, 9.5) {$\leftrightarrow$};
		\node [style=none] (101) at (20.25, 9.5) {};
		\node [style=none] (102) at (27.25, 11.5) {};
		\node [style=none] (103) at (27.25, 7.5) {};
		\node [style=none] (104) at (14.25, 7.5) {};
		\node [style=none] (105) at (14.25, 11.5) {};
		\node [style=none] (106) at (19.5, 11.75) {\scriptsize{} 1};
		\node [style=none] (112) at (19.5, 7.75) {\scriptsize{} 3};
		\node [style=none] (114) at (19.5, 9) {\scriptsize{} 3};
		\node [style=none] (115) at (17, 7.75) {\scriptsize{} 3};
		\node [style=none] (118) at (17, 11.75) {};
		\node [style=none] (119) at (17, 11.75) {\scriptsize{} 1};
		\node [style=none] (124) at (22.5, 11.75) {\scriptsize{} 2};
		\node [style=none] (125) at (22.5, 11.15) {\scriptsize{} 2};
		\node [style=none] (127) at (22.5, 7.75) {\scriptsize{} 4};
		\node [style=none] (131) at (26.75, 7.75) {};
		\node [style=none] (132) at (24.75, 2.75) {\scriptsize{} 2};
		\node [style=none] (133) at (24.75, -1.25) {\scriptsize{} 4};
		\node [style=none] (134) at (25, 11.75) {\scriptsize{} 2};
		\node [style=none] (135) at (25, 7.75) {\scriptsize{} 4};
		\node [style=none] (136) at (13, 0.25) {$\underset{\text{Stack rule}}{\rightsquigarrow}$ };
		\node [style=none] (137) at (15.25, 13.5) {};
		\node [style=none] (138) at (26.25, 13.5) {};
		\node [style=none] (139) at (26.25, 5.5) {};
		\node [style=none] (140) at (15.25, 5.5) {};
		\node [style=none] (141) at (15.25, 4.5) {};
		\node [style=none] (142) at (26.25, 4.5) {};
		\node [style=none] (143) at (26.25, -3.5) {};
		\node [style=none] (144) at (15.25, -3.5) {};
		\node [style=none] (145) at (17, -5.5) {};
		\node [style=white dot] (146) at (18, -6.5) {1};
		\node [style=none] (147) at (19, -5.5) {};
		\node [style=none] (148) at (19, -7.5) {};
		\node [style=none] (149) at (17, -7.5) {};
		\node [style=none] (150) at (17, -9.5) {};
		\node [style=none] (151) at (19, -9.5) {};
		\node [style=none] (152) at (17, -11.5) {};
		\node [style=none] (153) at (19, -11.5) {};
		\node [style=gray dot] (154) at (18, -10.5) {3};
		\node [style=none] (155) at (20, -5) {};
		\node [style=none] (156) at (20, -12) {};
		\node [style=none] (157) at (16, -12) {};
		\node [style=none] (158) at (16, -5) {};
		\node [style=none] (159) at (18, -8.5) {$\updownarrow$};
		\node [style=none] (160) at (22.5, -5.5) {};
		\node [style=gray dot] (161) at (23.5, -6.5) {2};
		\node [style=none] (162) at (24.5, -5.5) {};
		\node [style=none] (163) at (24.5, -7.5) {};
		\node [style=none] (164) at (22.5, -7.5) {};
		\node [style=none] (165) at (22.5, -9.5) {};
		\node [style=none] (166) at (24.5, -9.5) {};
		\node [style=none] (167) at (22.5, -11.5) {};
		\node [style=none] (168) at (24.5, -11.5) {};
		\node [style=white dot] (169) at (23.5, -10.5) {4};
		\node [style=none] (170) at (25.5, -5) {};
		\node [style=none] (171) at (25.5, -12) {};
		\node [style=none] (172) at (21.5, -12) {};
		\node [style=none] (173) at (21.5, -5) {};
		\node [style=none] (174) at (23.5, -8.5) {$\updownarrow$};
		\node [style=none] (175) at (25.25, -9.5) {};
		\node [style=none] (176) at (20.75, -8.5) {$\leftrightarrow$};
		\node [style=none] (178) at (20, -8.5) {};
		\node [style=none] (179) at (27.5, -6.5) {};
		\node [style=none] (180) at (27.5, -10.5) {};
		\node [style=none] (181) at (14, -10.5) {};
		\node [style=none] (182) at (14, -6.5) {};
		\node [style=none] (183) at (19.25, -6.25) {\scriptsize{} 1};
		\node [style=none] (185) at (20.25, -6.25) {\scriptsize{} 1};
		\node [style=none] (186) at (20.25, -7.75) {\scriptsize{} 3};
		\node [style=none] (188) at (20.25, -10.25) {\scriptsize{} 3};
		\node [style=none] (189) at (19.25, -10.25) {\scriptsize{} 3};
		\node [style=none] (191) at (19.25, -9) {\scriptsize{} 3};
		\node [style=none] (192) at (16.75, -10.25) {\scriptsize{} 3};
		\node [style=none] (193) at (15.75, -10.25) {\scriptsize{} 3};
		\node [style=none] (194) at (15.75, -6.25) {\scriptsize{} 1};
		\node [style=none] (195) at (16.75, -6.25) {};
		\node [style=none] (196) at (16.75, -6.25) {\scriptsize{} 1};
		\node [style=none] (197) at (21.25, -6.25) {\scriptsize{} 2};
		\node [style=none] (198) at (21.25, -7.25) {\scriptsize{} 2};
		\node [style=none] (200) at (21.25, -10.25) {\scriptsize{} 4};
		\node [style=none] (201) at (22.25, -6.25) {\scriptsize{} 2};
		\node [style=none] (202) at (22.25, -6.85) {\scriptsize{} 2};
		\node [style=none] (204) at (22.25, -10.25) {\scriptsize{} 4};
		\node [style=none] (207) at (25.75, -6.25) {\scriptsize{} 2};
		\node [style=none] (209) at (25.75, -10.25) {\scriptsize{} 4};
		\node [style=none] (210) at (24.75, -6.25) {\scriptsize{} 2};
		\node [style=none] (211) at (24.75, -10.25) {\scriptsize{} 4};
		\node [style=none] (212) at (13, -8.75) {$\underset{\text{Compose rule}}{\rightsquigarrow}$ };
		\node [style=none] (213) at (15.25, -4.5) {};
		\node [style=none] (214) at (26.25, -4.5) {};
		\node [style=none] (215) at (26.25, -12.5) {};
		\node [style=none] (216) at (15.25, -12.5) {};
		\node [style=none] (217) at (15, -10.25) {\scriptsize{} 3};
		\node [style=none] (218) at (15, -6.25) {\scriptsize{} 1};
		\node [style=none] (219) at (26.5, -6.25) {\scriptsize{} 2};
		\node [style=none] (220) at (26.5, -10.25) {\scriptsize{} 4};
	\end{pgfonlayer}
	\begin{pgfonlayer}{edgelayer}
		\draw (13.center) to (10.center);
		\draw (10.center) to (11.center);
		\draw (11.center) to (12.center);
		\draw (12.center) to (13.center);
		\draw (4.center) to (3.center);
		\draw (3.center) to (2.center);
		\draw (2.center) to (0.center);
		\draw (0.center) to (4.center);
		\draw (5.center) to (7.center);
		\draw (7.center) to (8.center);
		\draw (8.center) to (6.center);
		\draw (6.center) to (5.center);
		\draw (30.center) to (27.center);
		\draw (27.center) to (28.center);
		\draw (28.center) to (29.center);
		\draw (29.center) to (30.center);
		\draw (21.center) to (20.center);
		\draw (20.center) to (19.center);
		\draw (19.center) to (17.center);
		\draw (17.center) to (21.center);
		\draw (22.center) to (24.center);
		\draw (24.center) to (25.center);
		\draw (25.center) to (23.center);
		\draw (23.center) to (22.center);
		\draw (1) to (18);
		\draw [in=-135, out=15, looseness=0.50] (9) to (36.center);
		\draw (9) to (26);
		\draw (18) to (37.center);
		\draw (26) to (38.center);
		\draw (40.center) to (1);
		\draw (39.center) to (9);
		\draw (81.center) to (78.center);
		\draw (78.center) to (79.center);
		\draw (79.center) to (80.center);
		\draw (80.center) to (81.center);
		\draw (72.center) to (71.center);
		\draw (71.center) to (70.center);
		\draw (70.center) to (68.center);
		\draw (68.center) to (72.center);
		\draw (73.center) to (75.center);
		\draw (75.center) to (76.center);
		\draw (76.center) to (74.center);
		\draw (74.center) to (73.center);
		\draw (96.center) to (93.center);
		\draw (93.center) to (94.center);
		\draw (94.center) to (95.center);
		\draw (95.center) to (96.center);
		\draw (87.center) to (86.center);
		\draw (86.center) to (85.center);
		\draw (85.center) to (83.center);
		\draw (83.center) to (87.center);
		\draw (88.center) to (90.center);
		\draw (90.center) to (91.center);
		\draw (91.center) to (89.center);
		\draw (89.center) to (88.center);
		\draw (69) to (84);
		\draw [in=-135, out=15, looseness=0.50] (77) to (101.center);
		\draw (77) to (92);
		\draw (84) to (102.center);
		\draw (92) to (103.center);
		\draw (105.center) to (69);
		\draw (104.center) to (77);
		\draw (137.center) to (138.center);
		\draw (138.center) to (139.center);
		\draw (139.center) to (140.center);
		\draw (140.center) to (137.center);
		\draw (141.center) to (142.center);
		\draw (142.center) to (143.center);
		\draw (143.center) to (144.center);
		\draw (144.center) to (141.center);
		\draw (158.center) to (155.center);
		\draw (155.center) to (156.center);
		\draw (156.center) to (157.center);
		\draw (157.center) to (158.center);
		\draw (149.center) to (148.center);
		\draw (148.center) to (147.center);
		\draw (147.center) to (145.center);
		\draw (145.center) to (149.center);
		\draw (150.center) to (152.center);
		\draw (152.center) to (153.center);
		\draw (153.center) to (151.center);
		\draw (151.center) to (150.center);
		\draw (173.center) to (170.center);
		\draw (170.center) to (171.center);
		\draw (171.center) to (172.center);
		\draw (172.center) to (173.center);
		\draw (164.center) to (163.center);
		\draw (163.center) to (162.center);
		\draw (162.center) to (160.center);
		\draw (160.center) to (164.center);
		\draw (165.center) to (167.center);
		\draw (167.center) to (168.center);
		\draw (168.center) to (166.center);
		\draw (166.center) to (165.center);
		\draw (146) to (161);
		\draw [in=-135, out=15, looseness=0.50] (154) to (178.center);
		\draw (154) to (169);
		\draw (161) to (179.center);
		\draw (169) to (180.center);
		\draw (182.center) to (146);
		\draw (181.center) to (154);
		\draw (213.center) to (214.center);
		\draw (214.center) to (215.center);
		\draw (215.center) to (216.center);
		\draw (216.center) to (213.center);
		\draw [in=-150, out=45, looseness=0.75] (101.center) to (84);
		\draw [in=-150, out=45, looseness=0.75] (36.center) to (18);
		\draw [in=-150, out=45, looseness=0.75] (178.center) to (161);
	\end{pgfonlayer}
\end{tikzpicture}

%% file: sections/conclusion.tex
\section{Conclusion}

\VyZX is a formal verification framework for the ZX-calculus.
Creating \VyZX came with several challenges that are unique to verifying a graphical language.
Finding a way to easily assign semantics to a graph is inherently difficult due to traditional graph structures not being idiomatic in proof assistants.
Block representation provides an inductive description of a graph that allows for easy proof while preserving the expressiveness that graphs provide.
As block representation made it challenging to write diagrams, we developed a graph representation that can act as a way to implement programs over ZX diagrams.
With these two views in place, we proved several core ZX diagram equivalences and added circuit ingestion from \SQIR.
With all these tools in place, we believe \VyZX has a future as a basis for writing programs over ZX diagrams.
In particular, our next step is to build a verified circuit optimizer in the style of \pyZX.
The core of \VyZX will continue to be improved as we approach new problems in implementing programs such as a verified circuit simulator.
We are confident that with \VyZX's evolution, it will provide a robust foundation for future work on the verification of graphical quantum calculi and their applications.